# From Classical Superposition of Waves to Quantum Interference: Three Level Quantum System for Two Entangled Photons

**Amir Djalalian-Assl**

51 Golf View Drive, Craigieburn, VIC 3064, Australia
Correspondence: amir.djalalian@gmail.com

**Abstract:** Properties and applications of a plasmonic cross-shaped nano-antenna is presented and compared to those of array of holes. A simple analytical model based on the superposition of waves are proposed and compared to the numerical results. A direct consequence of unequal path for two orthogonal surface waves leads to a coherent quantum interferometer with interesting properties. Mechanism behind the rotating surface charge densities and consequently, the formation of rotating resultant dipole moments is identified and the concept of Dipole-SPP-LSP-Stokes coupling is introduced. All of which leads to the most significant findings (a) foundation of a three-level quantum system for entangled photons, based on the polarization states of the transmitted light and (b) further encapsulation of the three-level quantum system into a continuous orthonormal set of pure quantum states (c) an inference on the nonexistence of spin in a single photon, hence the requirement of two photons to create such spin states.

## 1. Introduction

A single photon source emits energy in the form of one quantized unit of light at a time. Controlling the polarization state of a Single Photon Source (SPS) provides a mechanism for defining the computational basis states [1-3]. A more elaborated account on the photon statistics related to the bunched, coherence, anti-bunched, single photon states, entangled photons and second order correlation function $g^2(0)$ can be found in [4,5]. The driving force behind the development of SPSs and single photon detectors is mainly the quantum information science, including cryptography [6]. Quantum cryptography based on Bell's theorem was first outlined in [7], where the polarized photons were proposed as a replacement for the ½-spin particle interactions. The quantum mechanics and the algorithms behind the cryptography are elaborated on in [8] and beyond the scope of this report. Suffice to say that the two requirements, i.e. individual quanta and the entangled states could be satisfied by the polarization states of photons. Furthermore, it was argued that the vertical or the horizontal polarization states can only be defined relative to the emitter's and/or detector's position and orientation, hence not suitable for real-life applications. Therefore, to form the computational basis states, the left-handed and the right-handed circular polarization are more suitable [2]. However, to generate circularly polarized light (CPL) at least two photons are needed. For a brief background on circularly polarized light, ellipticity, amplitude and the phase requirements of the two constituting orthogonal modes see chapter 2 of my master thesis [9]. Basically, to produce CPL, there must exist two orthogonal optical sources that satisfy two conditions: **(A)** The phase difference between the two must be ±90°, and **(B)** The two modes must be equal in amplitudes. I have previously reported on plasmonic devices as a possible approach to achieve this . This report is divided into two main headings, namely "*Asymmetric Cross-shaped Nano-antenna*", "*Biperiodic Array of Circular Holes*" and "*An Argument on Photon Spin*" organized in a chronological order, describing gradual development of ideas, notions and conclusions, covering topics such as the resultant dipole moments, superposition of waves between two holes, array of coherent quantum interferometers, three level quantum system, entangled photons and photon spin. Some findings were based on initial hypothesis subject to careful design, but some were based on serendipitous observations of results during the tasks related to the former and some at later stages, which I would discuss and highlight under subheadings in each section.

## 1. Asymmetric Cross-shaped Nano-antenna

An *Oscillating* electric dipole moment is defined by $\boldsymbol{\mu}_e(t) = \boldsymbol{\mu} e^{-i\omega_0 t} = Q\mathbf{d}e^{-i\omega_0 t}$ [10], where $Q$ is the charge and $\mathbf{d}$ is the vector distance from $-Q$ to $+Q$, that defines the dipole axis by a unit vector $\hat{\mathbf{d}} = \boldsymbol{\mu}_e / \|\boldsymbol{\mu}_e\| = \mathbf{d}/\|\mathbf{d}\|$, where double bars signify the magnitude of the vectors. Here, $\omega_0 = c/\lambda_0$ is the frequency and $\lambda_0$ the wavelength

in vacuum. Oscillations of charge densities along the two armlengths of a cross-shaped nano-antenna are of Localized Surface Plasmons (LSP) by nature and may be considered as the resultant (or vector sum) of the two orthogonal dipole moments, $\boldsymbol{\mu}_x = L_x Q_x \hat{\mathbf{x}}$, $\boldsymbol{\mu}_y = L_y Q_y \hat{\mathbf{y}}$. Quantities $L_{x,y} Q_{x,y}$ are the magnitude of electric dipole associated with induced charges $Q_{x,y}$ at the tips of the armlengths with $L_{x,y}$. Considering a linearly polarized light $\mathbf{E}_i = \left(E_x \hat{\mathbf{x}} + E_y \hat{\mathbf{y}}\right) e^{-i\omega_0 t}$ with a polarizations angle $\alpha = \tan^{-1}\left[E_y / E_x\right]$ impinging on the cross at normal incidence, induced charges are $Q_x \propto E_0 A_x \cos(\alpha)$ and $Q_y \propto E_0 A_y \sin(\alpha)$, where $E_0 = \sqrt{E_x^2 + E_y^2}$. Here, $A_{x,y}$ are factors encapsulating all other intrinsic physical effects, that contribute to the suppression/excitation of the modes. Consequently $\boldsymbol{\mu}_x = E_0 L_x A_x \cos(\alpha) \hat{\mathbf{x}}$ and $\boldsymbol{\mu}_y = E_0 L_y A_y \sin(\alpha) \hat{\mathbf{y}}$. In other words, the cross and all its intrinsic physical properties operate on the incident field to produce its own dipole moments:

$$\begin{bmatrix} L_x A_x & 0 & 0 \\ 0 & L_y A_y & 0 \\ 0 & 0 & 1 \end{bmatrix} \begin{bmatrix} E_0 \cos(\alpha) \\ E_0 \sin(\alpha) \\ 0 \end{bmatrix} = \frac{1}{\sqrt{2}} \begin{bmatrix} \boldsymbol{\mu}_x e^{-i\omega_0 t} \\ \boldsymbol{\mu}_y e^{-i(\omega_0 t \pm \Phi)} \\ 0 \end{bmatrix} \qquad (1)$$

The concept of virtual electric and magnetic dipoles of an elliptical aperture in a metallic thin film was reported by Zakharian and Mansuripur [11,12]. They demonstrated two distinct virtual dipole orientations inside the aperture for two distinct incident polarizations, that is parallel and normal to the ellipse's major axis. I have also shown that a cross-shaped aperture in a silver film possesses a well-defined virtual dipole moment [13] that interacts with the nearby quantum emitter. Therefore, the analysis on dipolar activities of a cross-shaped nano-particle may also be applied to virtual dipoles with $\boldsymbol{\mu}_x = E_0 L_x A_x \sin(\alpha) \hat{\mathbf{x}}$ and $\boldsymbol{\mu}_y = E_0 L_y A_y \cos(\alpha) \hat{\mathbf{y}}$. The resultant dipole moment being the RHS of equation (1) is of interest here and may be written as:

$$\boldsymbol{\mu}_e(t) = \left(\boldsymbol{\mu}_x e^{-i\omega_0 t} + \boldsymbol{\mu}_y e^{-i(\omega_0 t \pm \Phi)}\right) \Big/ \sqrt{2} \qquad (2)$$

That is the superposition of the two orthogonal dipole states. Here, the phase difference, Φ, is to cater for $L_x \neq L_y$. However, in contrary to the classical dipole where **d** has a physical meaning, the resultant dipole and its axis in this case are purely virtual and dependent on all the factors mentioned above, hence in some cases $\mathbf{d}/\|\mathbf{d}\|$ is meaningless. Instead the unit vector for the dipole axis must be calculated using $\hat{\mathbf{d}} = \boldsymbol{\mu}_e / \|\boldsymbol{\mu}_e\|$ which is valid in all cases. As an example, consider a simple case of a *symmetric* cross with $L_x = L_y$ and $A_x = A_y$, where $\alpha = 45°$ leads to $\boldsymbol{\mu}_x = \boldsymbol{\mu}_y$ and $\Phi = 0$. In this case, the direction of the resultant dipole and the unit vector defining the its axis is given by:

$$\hat{\mathbf{d}}_{lin} = \left(\boldsymbol{\mu}_x + \boldsymbol{\mu}_y\right) \Big/ \left\|\boldsymbol{\mu}_x + \boldsymbol{\mu}_y\right\| = (\hat{\mathbf{x}} + \hat{\mathbf{y}}) \Big/ \sqrt{2} \qquad (3)$$

and when acted on by the oscillating term $e^{-i\omega t}$, it would experience a time harmonic change in direction along the {-135°,45°} line with respect to the *x*-axis. Let us denote equation (3) as a *linear unit vector* hence the subscript "*lin*". In fact, for a symmetric cross, $\hat{\mathbf{d}}_{lin}$ aligns itself with α for all values of α and $\lambda_0$, with the far-field radiation pattern being that of a classical oscillating dipole, that is toroidal in shape, where the field intensity dropping to zero along $\hat{\mathbf{d}}_{lin}$. These were confirmed numerically. For a typical radiation pattern of an oscillating dipole see [10]. Now, consider another simple scenario with respect to an *asymmetric* cross. Let us assume that for some α and $\lambda_0$, conditions (**A**) and (**B**) are satisfied, leading to $\boldsymbol{\mu}_x = \boldsymbol{\mu}_y$ and $\Phi = 90°$. In this case:

$$\hat{\mathbf{d}}_{cir} = \left(\boldsymbol{\mu}_x + \boldsymbol{\mu}_y e^{-i(\pi/2)}\right) \Big/ \left\|\boldsymbol{\mu}_x + \boldsymbol{\mu}_y e^{-i(\pi/2)}\right\| = (\hat{\mathbf{x}} - i\hat{\mathbf{y}}) \Big/ \sqrt{2} \qquad (4)$$

that is complex phasor when operated on by the oscillating term $e^{-i\omega_0 t}$, hence $\hat{\mathbf{d}}_{cir}$ not being confined to any linear direction in the *x*-*y* plane (i.e. the antenna plane), but instead rotating about the *z*-axis (i.e. the optical axis) while maintaining its unit magnitude at all time. For a lack of a better term, let us call it a *circular unit vector*.

I presented the asymmetric cross-shaped nano-antenna and the concept of resultant dipole moment in a poster session [14] with relevant figures included here, see also section 4.3 of my thesis arXiv:1912.10888 [physics.optics]. I have described the relevant aspects of the numerical modelling in [15]. The only difference here is the use of asymmetric copper cross rather than the gold nano-rods. Spectra in Figure 1(a) shows the numerical results of an asymmetric cross-shaped nano-antenna, vs the wavelength when excited by a normally incident linearly polarized light in the range 0° ⩽ α ⩽ 90°, where α is the angle of polarization . Radar Cross-Section, $RCS \propto |E_{\text{far}}|^2 / |E_{\text{i}}|^2$ , was calculated in the *x-y* plane from the scattered far-field $E_{\text{far}}$ and the background field $E_{\text{i}}$. The two orthogonal modes at $\lambda_{\text{res1}} = 820$ nm and $\lambda_{\text{res2}} = 890$ nm are associated with $L_{\text{x}} = 80$ nm and $L_{\text{y}} = 95$ nm respectively. To excite the two modes equally, α was set to 32° from the *x*-axis. Naturally, due the separation of the two resonances, Φ ≠ 0. Figure 1(b) shows the far-field radiation patterns with α = 32° in the range of $\lambda_{\text{res1}} \leqslant \lambda_0 \leqslant \lambda_{\text{res2}}$.

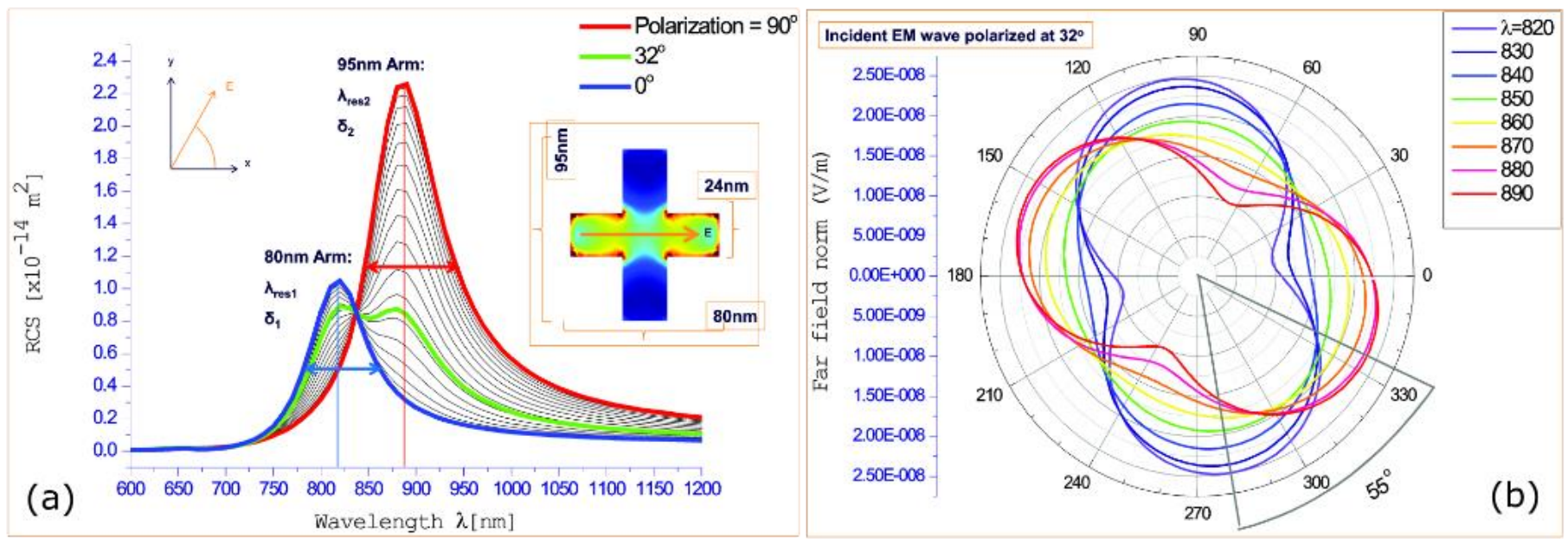

Figure 1: (a) Radar cross-section, $RCS \propto |E_{\text{far}}|^2 / |E_{\text{i}}|^2$ , calculated in the *x-y* plane from the scattered far-field $E_{\text{far}}$ and the background field $E_{\text{i}}$.(b) Far-field radiation patterns for $\lambda_{\text{res1}} \leqslant \lambda_0 \leqslant \lambda_{\text{res2}}$ in the *x-y* plane with α = 32°, when the two orthogonal modes are equally excited.

The angular changes in radiations vs the incident wavelength were comparable to those of the gold "T" antenna [15]. But unlike the "T" antenna, far-field radiation patterns of an asymmetric cross showed a 180° rotational symmetry. More importantly, the radiation pattern associated with $\lambda_0 = 850$ nm was close to a perfect circle. That was an evidence of perpetual rotation of the resultant dipole moment about the *z*-axis driven only by the time harmonic term $e^{-i\omega_0 t}$ . In other words, the dipole axis was defined purely by $\hat{\mathbf{d}}_{cir}$ . For other values of $\lambda_0$, radiation patterns were elliptic, never dropping to zero, suggesting superpositions of two kinds of dipolar activities, hence $\hat{\mathbf{d}} = \left(a\hat{\mathbf{d}}_{lin} + b\hat{\mathbf{d}}_{cir}\right) / \left\| a\hat{\mathbf{d}}_{lin} + b\hat{\mathbf{d}}_{cir} \right\|$, where *a*, *b* signifies the strength of each type of dipole in the sum.

I reported a similar concept with respect to an asymmetric cross-shaped aperture in a bullseye (BE) setting in a silver screen [16] which revealed the correlation between surface effects and the transmitted state of polarization. The model consisted of a cross-shaped aperture with $L_{\text{x}} = 150$ nm and $L_{\text{y}} = 220$ nm at the center of a BE structure with concentric circular corrugations having an inner radius, $r_{\text{in}} = 710$ nm and a period $P = 650$ nm. Dimensions were optimized for $\lambda_0 = 700$ nm. Device was illuminated by a normally incident linearly polarized light at $\lambda_0 = 700$ nm from the glass substrate, and the state of polarization was calculated from the transmitted field. Figure 2(a) shows the numerically calculated Stokes parameters obtained from the transmitted field vs α. Here $S_1$, $S_2$ and $S_3$ range from -1 to 1, signifying degrees of vertical/horizontal, diagonal and circular polarizations respectively. Figure 2(b) depicts the surface charge densities calculated at α = 90° where the transmitted field showed $S_3 = S_2 = 0$ and $S_1 = -1$. Figure 2(c) represents the spiral surface charge densities launched by the cross when α = 46° corresponding to $S_1 = S_2 = 0$ and $S_3 = 1$. Consequently, a clear link between the transmitted state of polarization and the surface effects was possible by inference. See Appendix A, for method of calculating the surface charge density.

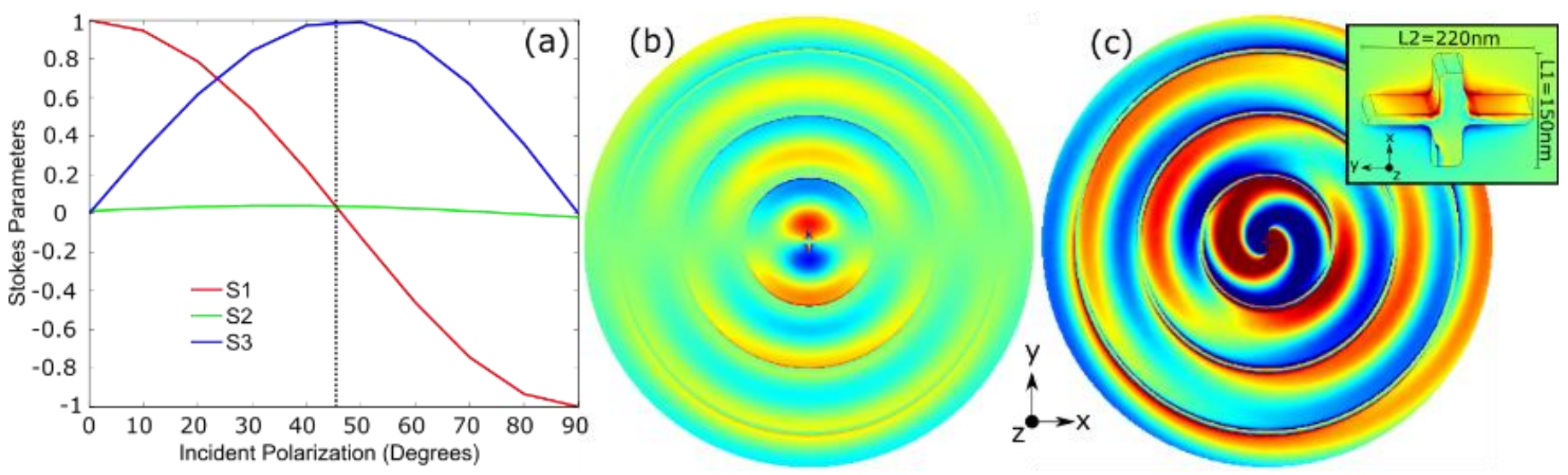

Figure 2: (a) Calculated Stokes parameters obtained from the transmitted field through an isolated cross-shaped aperture with $L_x = 220$ nm and $L_y = 150$ nm, vs α. (b) Surface charge densities calculated at α = 90° related to the transmitted state of polarization $S_3 = S_2 = 0$ and $S_1 = -1$. (c) Spiral surface charge densities produced when $\alpha = 46°$, with the transmitted state of polarization $S_1 = S_2 = 0$ and $S_3 = 1$. Inset shows LSPs inside the aperture.

## *2.1 Significant Observations and Applications*

### 2.1.1 Applications in Radiofrequency Antennas

In general, there are strong analogies between optical and radiofrequency (RF) dipole antennas. One can hypothesis on a pair orthogonal RF dipole antenna satisfying conditions **(A)** and **(B).** Consider the normal vector $\mathbf{n_r}$ defining the plane of the resultant dipole moment such that $\boldsymbol{\mu}_x \perp \boldsymbol{\mu}_y \perp \mathbf{n_r}$. From Figure 2(c), one can intuitively infer that the radiation from such a pair adheres to $\mathbf{E} \perp \mathbf{n_r}$ for all $\mathbf{k}_0 \perp \mathbf{n_r}$, eliminating the need for mechanically driven revolving RF dipole antennas. Here $\mathbf{k}_0$ is the wavevector defining the direction of propagation from the resultant dipole to the point of observation and $\mathbf{E}$ is the electric field vector. Naturally, when considering the transmission of such a pair along $\mathbf{k}_0 \parallel \mathbf{n_r}$, a circularly polarized radio waves (CPRW) would be detected, and that may have applications in astronomy, just as CPL has applications in microscopy.

### 2.1.2 Dipole-SPP-LSP-Stokes coupling

Figure 2(b) and Figure 2(c) are clear signs of dipolar activities associated with $\hat{\mathbf{d}}_{lin}$ and $\hat{\mathbf{d}}_{cir}$. Correlation between the transmitted state of polarization and the surface activities such as SPPs, LSPs and $\hat{\mathbf{d}}$ was established, meaning that by measuring the transmitted state of polarization one could infer the state of SPP, LSP on the surface and ultimately $\hat{\mathbf{d}}$ inside the aperture. This dipole-SPP-LSP-Stokes coupling inferred that rotating surface waves may induce rotating resultant dipole moments inside symmetric apertures such as circular holes, as I will expand on in later sections.

### 2.1.3 Mechanism governing the spiral surface waves

The magnetic field associated with the resultant dipole moment above, though orthogonal to the electric field, rotates in the *x*-*y* plane and about the *z*-axis just as the electric field does. Hypothetically, it is possible for a constant magnetic field, $B_z$ in the *z*-direction, either applied externally or established by the hole, to induce cyclotronic electrons inside or in the vicinity of the hole that rotate about the *z*-axis. In a flat metallic slab the cyclotron resonance has a frequency [17]:

$$\omega_c = \frac{eB_z}{m^*} \qquad (5)$$

where $e$ is the charge and $m^*$ is the effective mass of the electron, hence clearly a function of $B_z$. Whereas my numerical results confirmed the rotation frequency of the spiral to be $\omega_0 = 2\pi/T$ where $\omega_0$ and $T$ are the frequency and the period of the incident wave respectively. Furthermore, my numerical analysis and experiments were carried out in the absence of any applied magnetic field. Consequently, at this stage, I tend to eliminate any possible scenarios where any cyclotronic electrons being formed inside or near the aperture when the device is exited with an incident light at $\lambda_0 = 700$ nm and $\alpha = 46°$. Nevertheless, a future study on plasmonic holes with constant magnetic field, $\pm B_z$ applied externally, while the device is excited with an incident light, would make an interesting research project. How does $\pm B_z$ impacts the SPPs? How the presence of a metallic hole impact the $\omega_c$? Alternatively, in the absence of any incident light, could such cyclotrons match the SPP's momentum, so to ignite a spiral or any other type of SPPs at some frequency? And if so, could this lead to any scattered light into the frees-pace.

Mechanism governing the spiral surface waves is clear and simple. I have reported on the SPP waves launched under the forced vibration by a virtual dipole of an aperture previously [13,18]. The only difference here is the formation of the resultant dipole moment and its spin angular momentum, when the device is exited with an incident light at $\lambda_0 = 700$ nm and $\alpha = 46°$. The angular frequency of the spiral about the *z*-axis is that of the drive, $\omega_0$, as I mentioned above. Meaning that the resultant dipole has a time dependent orientation, $\theta_{\text{dipole}}(t) = \omega_0 t$, in the *x*-*y* plane while continuously launching SPPs. Consequently, SPPs propagate radially away from the aperture with the wavevector, $\mathbf{k}_{\text{SPP}}$, in directions that are time lapsed, and as such there is no intrinsic orbital or spin angular moment involved in $\mathbf{k}_{\text{SPP}}$. The spiral appearance of the surface charge densities is purely due to the time lapsed between the consecutive points that are in phase yet conceived at different points in time. Any set of such points on surface waves, may be traced along a single strand of an Archimedean spiral line, such as $\frac{\lambda_{SPP}}{2\pi}\theta_{SPP}$, if expressed in polar coordinates. For a given phase, $\phi_j$, where $j$ is an integer, a single strand on the spiral SPP may be described in Cartesian coordinates:

$$\begin{bmatrix} x(t,\phi_i,\theta_{SPP}) \\ y(t,\phi_i,\theta_{SPP}) \end{bmatrix} = \begin{bmatrix} \mathrm{Cos}[\omega_0 t] & \mathrm{Sin}[\omega_0 t] \\ -\mathrm{Sin}[\omega_0 t] & \mathrm{Cos}[\omega_0 t] \end{bmatrix} \begin{bmatrix} \mathrm{Cos}[\phi_i] & \mathrm{Sin}[\phi_i] \\ -\mathrm{Sin}[\phi_i] & \mathrm{Cos}[\phi_i] \end{bmatrix} \begin{bmatrix} \dfrac{\lambda_{SPP}\theta_{SPP}\cos(\theta_{SPP})}{2\pi} \\ \dfrac{\lambda_{SPP}\theta_{SPP}\sin(\theta_{SPP})}{2\pi} \end{bmatrix} \quad (6)$$

where the first and the second term on the RHD are the rotation/transformation matrices catering for the time dependent rotation and the phase strand, and the third term is the Archimedean spiral vector catering for the special distribution of SPPs. Note that for simplicity, above equations were derived in the absence of any corrugation. To include the impact of the corrugations one must also incorporate the superposition of forward and reflected backward propagating SPPs as I reported in [13]. Nevertheless, the mechanism governing the spiral waves remain essentially the same.

*2.2 Experimental Demonstration*

I previously reported a number of spectra related to isolated cross-shaped apertures, perforated in a 100 nm silver film supported on a glass substrate [19]. Here I have included two additional spectra, $L$ = 110 nm and $L$ = 230 nm, with the full wavelength range as detected by the spectrometer, see Figure 3(a). To eliminate any impact due to possible disparities between the two armlengths, I used a linearly polarized light aligned with one arm of the cross. Discrepancies between the simulations and the experiment is obvious. Numerical results showed that spectra associated with $L_x$ = 150 nm and $L_y$ = 220 nm intercept at $\lambda_0$ = 700 nm, whereas the fabricated crosses with armlength $L$ = 150 nm and $L$ = 210 nm intercept at $\lambda_0$ = 660 nm. In fact, by visually inspecting the experimental results and based on the separation of resonances, I would infer that the two most candid armlengths that could satisfy conditions **(A)** and **(B)**, would have been $L_x$ = 110 nm (line in blue) and $L_y$ = 230 nm (line in red) with target wavelength $\lambda_0$ = 660 nm, but that meant moving away from my target wavelength. I can also infer, for example, that $L_x$ = 110 nm and $L_y$ = 150 nm could not satisfy conditions **(A)** and **(B)** fully given their excessive spectral overlap. And that raises doubts on some experimentally obtained values of Stokes parameters for an *array* of asymmetric crosses with inadequate separation between resonances[20]. In chapter 8 [21], I demonstrated that for silver film thicknesses less than 50 nm, incident field leaks and is transmitted through the film. What percent of the incident film was transmitted through the 40 nm film as reported in [20], and by how much impacted the Stokes parameters was not clarified.

Another undesirable effect observed in Figure 3(a), was the wider than expected FWHM associated with, for example $L$ = 150 nm (line in green), which directly impacts the phase requirements. The non-Lorentzian line-shape of $L$ = 150 nm spectrum with quasi double peaks around 640 $\leqslant \lambda_0 \leqslant$ 685 nm, and the appearance of two distinct resonances at $\lambda_0$ = 880 nm and $\lambda_0 \geqslant$ 1000 nm, also hint at superpositions of multiple resonances of various origins. Weaker resonances at $\lambda_0$ = {440 nm, 500 nm} which also appear for $L$ = 170 nm and $L$ = 110 nm respectively, may readily be attributed to higher order aperture modes and are not of concerns here. Notably, the left-leg of $L$ = 150 nm spectrum overlaps with that of $L$ = 110 nm, whereas the right-leg is red-shifted by ~45 nm. In the case of $L$ = 150 nm, if one attributes the mode at $\lambda_0 \approx$ 640 nm to aperture's thermal radiation (which, in the absence of an external heat source, depends on the intensity of the incident field and ohmic losses), the appearance of a quasi-peak at $\lambda_0$ = 685 nm may then be attributed to the fundamental mode associated with the virtual electric dipole of the aperture whose resonance is a function of the arm-length. In fact, the red-shift observed on the right-legs of the spectra are all consistent with increase in the arm-lengths. This implies that for $L$ = 110 nm and $L$ = 230 nm, either the thermal mode did not exist or it coincided with fundamental mode, hence their near-perfect Lorentzian peaks. Remaining peaks beyond the fundamental modes may then be explained based on Zakharian and Mansuripur [11,12] who reported on the formation of virtual magnetic dipole in an elliptical metallic aperture. Although magnetic and electric dipoles may coexist at the same wavelength, it is not unreasonable to assume that magnetic dipoles *resonate* at a different frequency to electric dipoles. Subsequently, I tend to believe that peaks at (for example) $\lambda_0$ = 970 nm, $\lambda_0$ = 880 nm, $\lambda_0 \geqslant$ 1000 nm and $\lambda_0$ = 956 nm, see Figure 3(b-d), are due to the resonating magnetic modes.

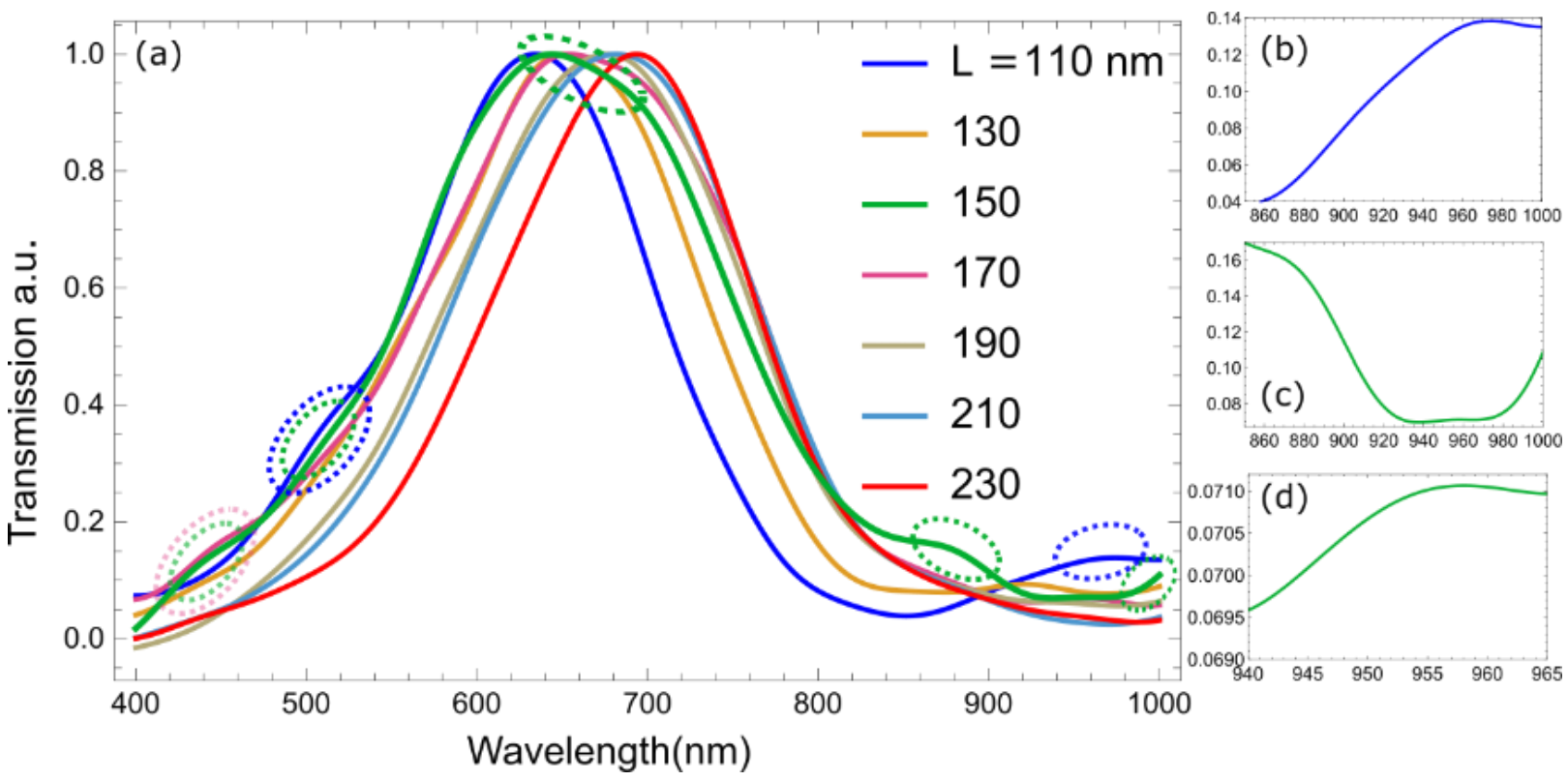

Figure 3: Optical response of single isolated symmetric cross-shaped apertures with various arm-lengths, perforated in a 100nm silver film supported on a glass substrate. Spectra was taken in dark-field mode normalized to the max.

In any case, based on the target dimensions obtained from the numerical models, and the spectra of the single crosses obtained experimentally, my first attempt to fabricate the bullseye device was less than satisfactory, compare the target dimensions above to the those in Figure 4(a). In summary, $L_x = 200$ nm, $L_y = 270$ nm, $707 \leqslant r_{in} \leqslant 738$ nm and $P = 630$ nm. Figure 4(b) depicts the lights transmitted through and scattered by the device when illuminated with an incandescent light. Clearly the gaps resonate at a shorter wavelength than the aperture. Moreover, spectral measurements seen in Figure 4(c), were produced using a confocal microscope being focused on the aperture, eliminating any possible radiation from the corrugations. And although it revealed the presence of the two sought orthogonal modes, it showed that one mode persisted for all incident polarizations. Thus, the optimum value for $\alpha$ could not be determined and it was set to 45°. Despite my previous suggestion and I quote: "*The peak at λ=715 nm which is present in all cases, is attributed to both the SPP Bloch mode associated to the periodic corrugations and the LSPR associated with the shorter arm of the asymmetric cross*.[16]", though a probable cause, one must not dismiss other factors such as the off-center position of the cross with respect to the center of the corrugations, hence yet another two modes associated with the inner circle. Given that $S_3$ parameter peaked at $\lambda_0 = 715$ nm, see Figure 4(d), one may even conclude that the two modes of the cross as well as that of the corrugations were located around 710 nm $\leqslant \lambda_0 \leqslant$720 nm, thus too close to be distinguished. But there is another possibility. Aperture's thermal radiation as explained above, may explain the persisting peak at $\lambda_0 = 715$ nm in Figure 4(c), and the two peaks at $\lambda_0 = 775$ nm and $\lambda_0 = 825$ nm in Figure 4(c)-line-in-blue may well be the magnetic modes. Having said all that, a higher degree of circularly polarized light could have been possible if the measurements were repeated with various values of $\alpha$, a technique I developed at later stages.

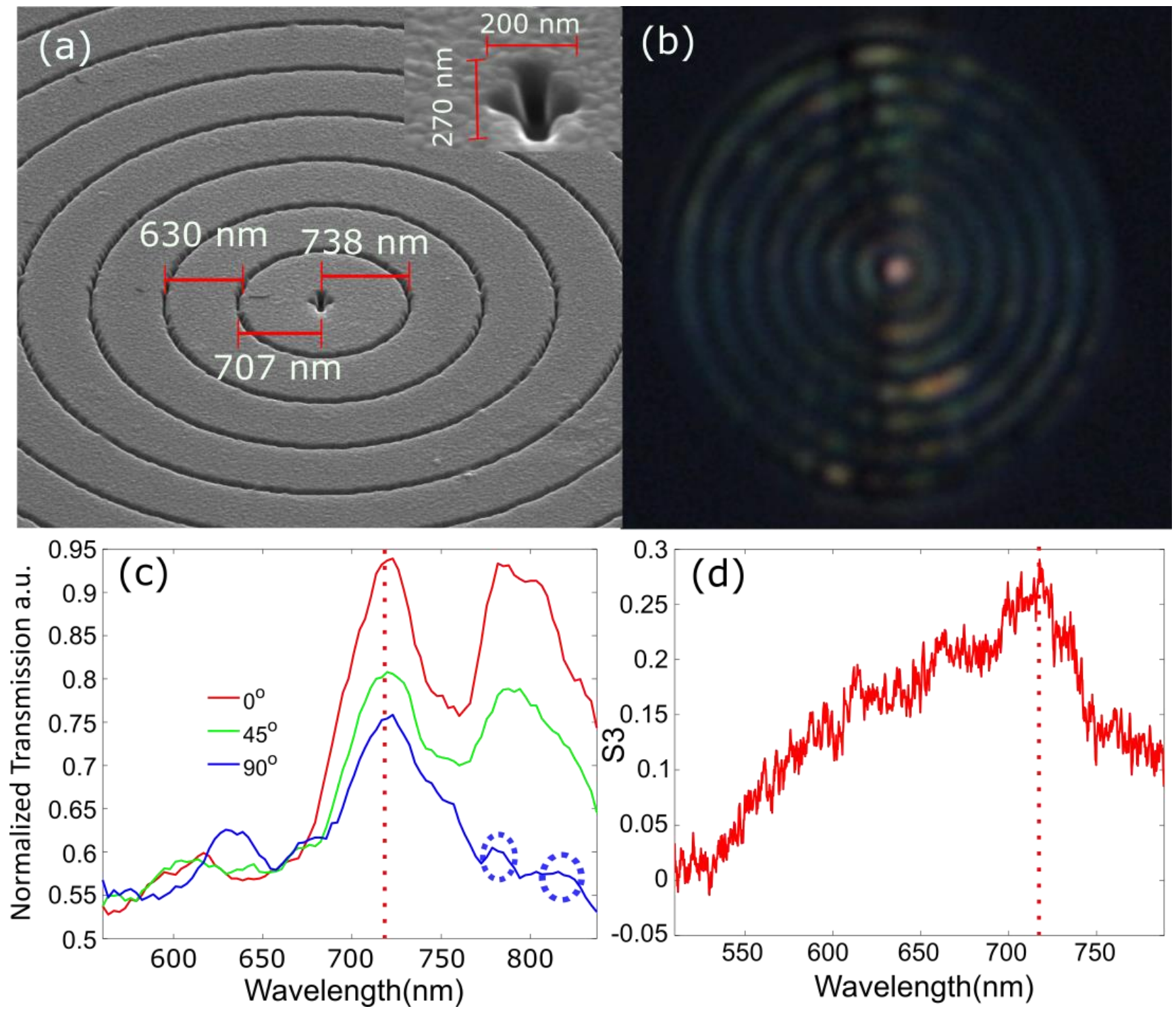

Figure 4: (a) SEM images of the fabricated BE structure with an asymmetric cross-shaped aperture. (b) Transmitted optical image taken using an inverted microscope when illuminated with a halogen light. (c) Normalized transmission through the device. (c) $S_3$ calculated from the transmitted field.[16]

In summary, despite promising numerical results the experimental data, being subject to fabrication errors, were less than impressive. But the concept was proven which led to an improvement at later stage, see chapter 9 of my thesis [21]. A simpler structure with fewer interlocked effects was the next logical step as described in the following section.

## 2. Biperiodic Array of Circular Holes

Drezet *et al.*[22] proposed a BE structure with periodic elliptical corrugations such that $a_n = b_n + \delta L$, where $\delta L$ is the length difference between the long axis, $a_n$, and the short axis, $b_n$, of the $n^{th}$ concentric ellipse. It was then implied that for the SPP Bloch waves to satisfy the condition to produce CPL, the length difference must simply satisfy $\delta L \times k_{SPP} = \pi/2$. However, as I will explain shortly, this approach is inappropriate, given the phase relation is calculated inaccurately and the strength of the two orthogonal surface waves are not considered. Considering a rectangular primitive lattice with constants $P_x$ and $P_y$, for such a plasmonic meta-surface to produce a CPL, condition **(A)** dictates $\Phi_{SPP,x} - \Phi_{SPP,y} = \pi/2$, where $\Phi_{SPP,x,y}$ represent the relative phases associated with SPPs propagating along the $P_x$ and $P_y$ directions respectively. Condition **(B)** requires the two orthogonal surface modes be equal in amplitudes to prevent ellipticity. These are the fundamental physical effects used in this report to explain other physical effects which are experimentally confirmed by means of measuring the polarization state of the transmitted light. It is customary among the plasmonic community to quote the following relations governing the surface wave in periodic structures:

$$P = 2\pi\sqrt{i^2 + j^2} \Big/ k_{SPP} \qquad (7)$$

where

$$k_{SPP} = 2\pi \,\mathrm{Re}\left[\sqrt{\varepsilon_m \varepsilon_d \big/ \left(\varepsilon_m + \varepsilon_d\right)}\right] \Big/ \lambda_0 \qquad (8)$$

Using equations (7) and (8), the period of a square array of holes that supports SPPs at its glass/silver interface at $\lambda_0 = 700$ nm was found to be $P = \lambda_{SPP} = 433$ nm. Applying Drezet's suggestion to the square array, the detuning in each direction should be given (erroneously) by $\Delta P = \pm (\pi/4)/k_{SPP} = \pm 54$ nm. However, in *any* periodically patterned surfaces, be it periodic concentric corrugations or hole arrays, one must not ignore the superposition of the forward propagating SPP with its own reflection by the scatterers. In the case of concentric surface corrugations, I have highlighted how the amplitude of a surface waves varies at a scattering point with respect to the scatterer spacings when one takes into account the superposition [13]. In the case of hole arrays, I have discussed the failure of the Bloch theorem in predicting the spectral peaks and proposed a model based on superpositions of surface waves see chapter 6 of my thesis [21]. Given the normal to the surface component of SPPs being an odd function with respect to the center of the hole [13], one may formulate their superposition at the center of the hole, $(x,y) = (0,0)$ as:

$$\Psi_z(k_{SPP}P_{x,y})\big|_{x,y=0} = \frac{1}{3}\left[\psi_1(0)+\psi_1\left(2k_{SPP}P_{x,y}\right)+\psi_2\left(\pi-k_{SPP}P_{x,y}\right)\right] \quad (9)$$

where $\psi_{1,2}(kx) = e^{ikx}$ define waves with a wavenumber, $k$, having travelled a distance $x$, from its source identified by the subscripts. For schematics see Figure 5(a). To satisfy condition **(A)**, one must calculate the square of the amplitude and the relative phase vs. $P$ and then obtain the detuning about the center wavelength, $P = \lambda_{SPP} = 433$ nm from the result, (see Figure 5). The two orthogonal lattice constants are then determined to be $P \pm \Delta P = 433 \pm 21$ nm, i.e. $P_x = 412$ nm and $P_y = 454$ nm.

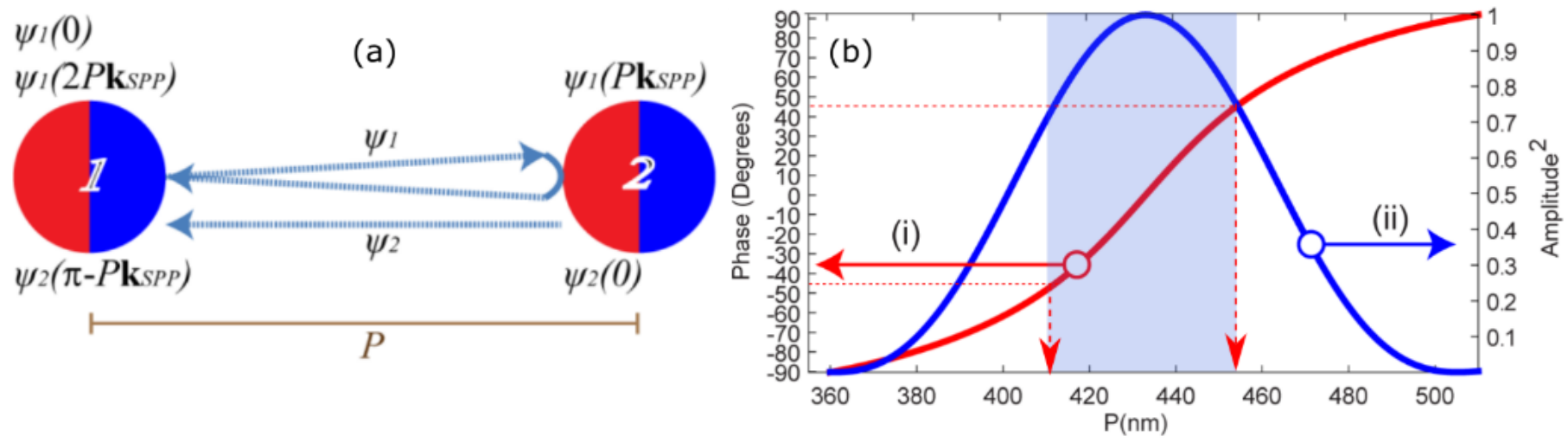

Figure 5: (a) Schematics for two holes interaction via the $z$-component of the SPP. (b) The square of the amplitude and the relative phase vs. the $P$, analytically calculated for $k_{SPP} = 2\pi/433$ (nm$^{-1}$) corresponding to $\lambda_0 = 700$ nm.

Parallel to the surface components of the SPPs, that are even functions with respect to the center of the hole, hence being responsible for the scattered power [13], may be expressed as:

$$\Psi_{x,y}(k_{SPP}P_{x,y})\big|_{x,y=0} = \frac{1}{3}\left[\psi_1(0)+\psi_1\left(2k_{SPP}P_{x,y}\right)+\psi_2\left(-k_{SPP}P_{x,y}\right)\right] \quad (10)$$

which also satisfies the condition **(B)**. Stokes parameters and the Degree of Polarization (DOP) for SPPs at the surface and about the lattice point $(x, y) = (0,0)$ may be calculated using:

$$\mathrm{S}_0\big|_{x=0} = \left|\Psi_z(k_{SPP}P_x)\right|^2 + \left|\Psi_z(k_{SPP}P_y)\right|^2 \quad (11)$$

$$\mathrm{S}_1\big|_{x,y=0} = -\left(\left|\Psi_z(k_{SPP}P_x)\right|^2 - \left|\Psi_z(k_{SPP}P_y)\right|^2\right)\Big/\mathrm{S}_0 \quad (12)$$

$$\mathrm{S}_2\big|_{x,y=0} = 2\,\mathrm{Re}\left[\Psi_z(k_{SPP}P_x)\Psi_z(k_{SPP}P_y)^*\right]\Big/\mathrm{S}_0 \quad (13)$$

$$\mathrm{S}_3\big|_{x,y=0} = -2\,\mathrm{Im}\left[\Psi_z(k_{SPP}P_x)\Psi_z(k_{SPP}P_y)^*\right]\Big/\mathrm{S}_0 \quad (14)$$

$$\mathrm{DOP} = \sqrt{\mathrm{S}_1^2+\mathrm{S}_2^2+\mathrm{S}_3^2} \quad (15)$$

I have also adapted the frequency response function $R(\omega) = \gamma^2\omega^2\Big/\left[\left(\omega_0^2-\omega^2\right)^2+\gamma^2\omega^2\right]$ [23], to suit the individual lattice modes $k_{i,j}$:

$$R_{i,j} = \gamma_{i,j}^2 k_{\mathrm{SPP}}^2\Big/\left[\left(k_{i,j}^2-k_{\mathrm{SPP}}^2\right)^2+\gamma_{i,j}^2 k_{\mathrm{SPP}}^2\right] \quad (16)$$

where $i$ and $j$ are integers, $\gamma_{i,j} = k_{i,j} / Q_{i,j}$, $k_{1,0} = 2\pi / P_x$, $k_{0,1} = 2\pi / P_y$, $k_{i,j}\big|_{i,j\neq0} = \sqrt{i^2 k_{1,0}^2 + j^2 k_{0,1}^2}$ and $Q_{i,j}$ is the quality factor for the $(i, j)$ mode. Figure 6(a) depicts $R_{i,j}$ with quality factors $Q_{0,1} = 5$, $Q_{1,1} = 6$ and $Q_{0,2} = 8$ chosen such that the full-width-half-max of the spectra produced by equation (16) match those produced by equation (10). Peak positions for the (0.1) and (0,2) modes obtained from equation (16), coincide precisely with those obtained from $\mathrm{Re}\left(\Psi_{x,y}\right)^2\Big|_{x,y=0}$ using equation (10), see Figure 6(b). Stokes parameters and the DOP calculated from equations (10)-(15) are depicted in Figure 6(c). The superposition also predicts resonances in the vicinity of $(1,1)_{\mathrm{glass}}$ mode, compare $R_{1,1}$ in Figure 6(a) to $\mathrm{Re}\left(\Psi_{x,y}\right)^2\Big|_{x,y=0}$ in Figure 6(b). Given that equation (10) concerns only two holes, that is a 1D array, resonances in the vicinity of $(1,1)_{\mathrm{glass}}$ are purely due to the superposition of surface waves and cannot be thought of as lattice modes. So, lets label them as quasi-$(1,1)_{\mathrm{glass}}$ modes for convenience.

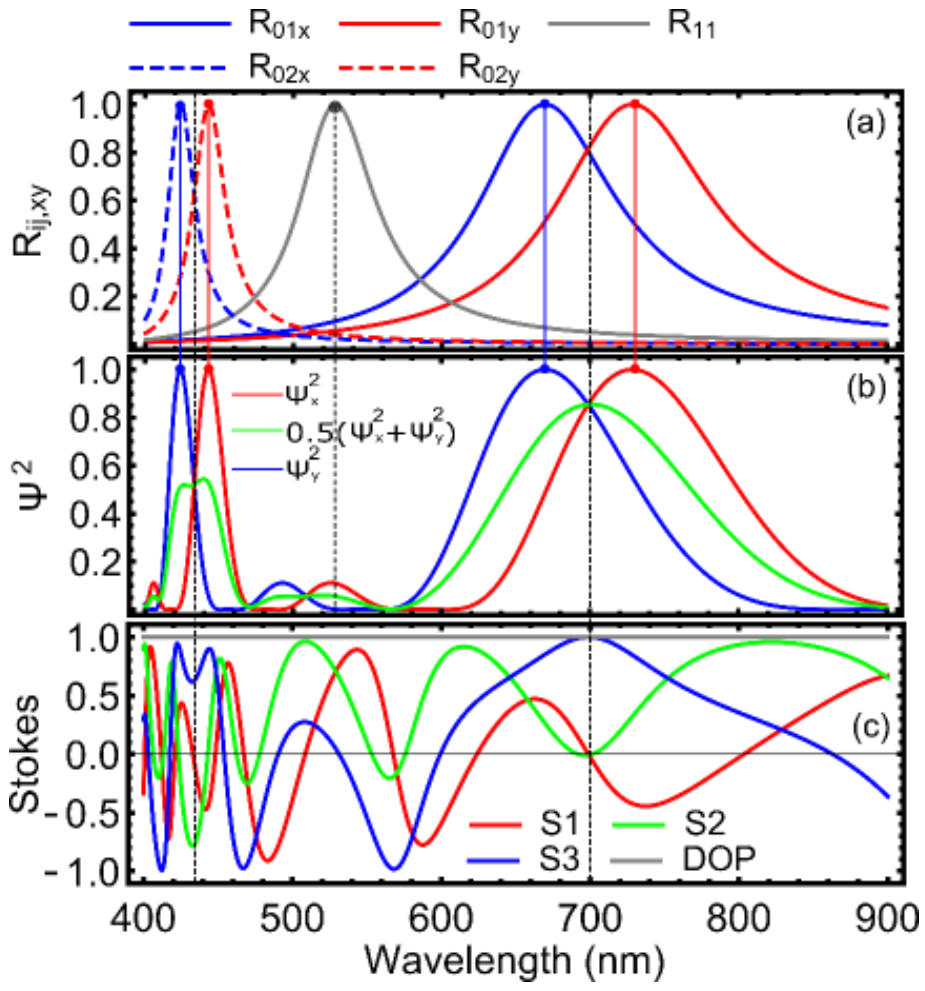

Figure 6: (a) $R_{i,j}(k_{\mathrm{SPP}})$, (b) $\mathrm{Re}\left(\Psi_{x,y}\right)^2\Big|_{x,y=0}$, (c) Stokes parameters and the DOP calculated from equations (10) -(15).

The model clearly predicts the peak resonances associated with a plasmonic empty lattice and explains the dumping $\gamma$. So far, the proposed model was based purely on of surface waves when considering a virtual lattice having no holes. Consequently, many other known and unknown effects relevant to subwavelength apertures are excluded. I have described the numerically modelled array of holes having diameters $d$ = 200 nm perforating a $h$ = 100 nm silver film supported on a glass substrate previously [21,24], whereby, a 3D model of a unit cell of the array was simulated using Finite Element Method (FEM) using COMSOL Multiphysics/RF module (EM Wave, Frequency Domain) with Stationary Solver. The unit cell consisted of glass/silver/air layers with top/bottom boundaries terminated with scattering boundary condition (SBC). Refractive index of the material filling the hole was set to $n_{\mathrm{h}}$ = 1. Side boundaries of the unit cell were configured with Periodic Boundary Condition (PBC). Silver film was set in the $x$-$y$ plane. Structure was illumination from the glass substrate by an incident wave propagating in the +z direction for a design wavelength of $\lambda_0$ = 700 nm under the normal incidence. The electric field was calculated at the top air-side boundary of the cell. The periodicity $P$ = 394 nm corresponded to the fundamental resonant mode $(1,0)_{\mathrm{glass}}$. A parametric sweep was performed over $P_y$, while keeping $P_x$ = 394 nm. Transmission and relative phase differences between the $x$ and the $y$ components of the transmitted electric field were calculated, see Figure 7. The two orthogonal lattice constants, $P_x \approx 368$ nm and $P_y \approx 407$ nm satisfied the phase difference of 90°, hence satisfying condition **(A).** Note that the total detuning $P_x$ - $P_y$ = 39 nm obtained from the simulation is close to the analytical value $2\Delta P$ = 42 nm using equation (9).

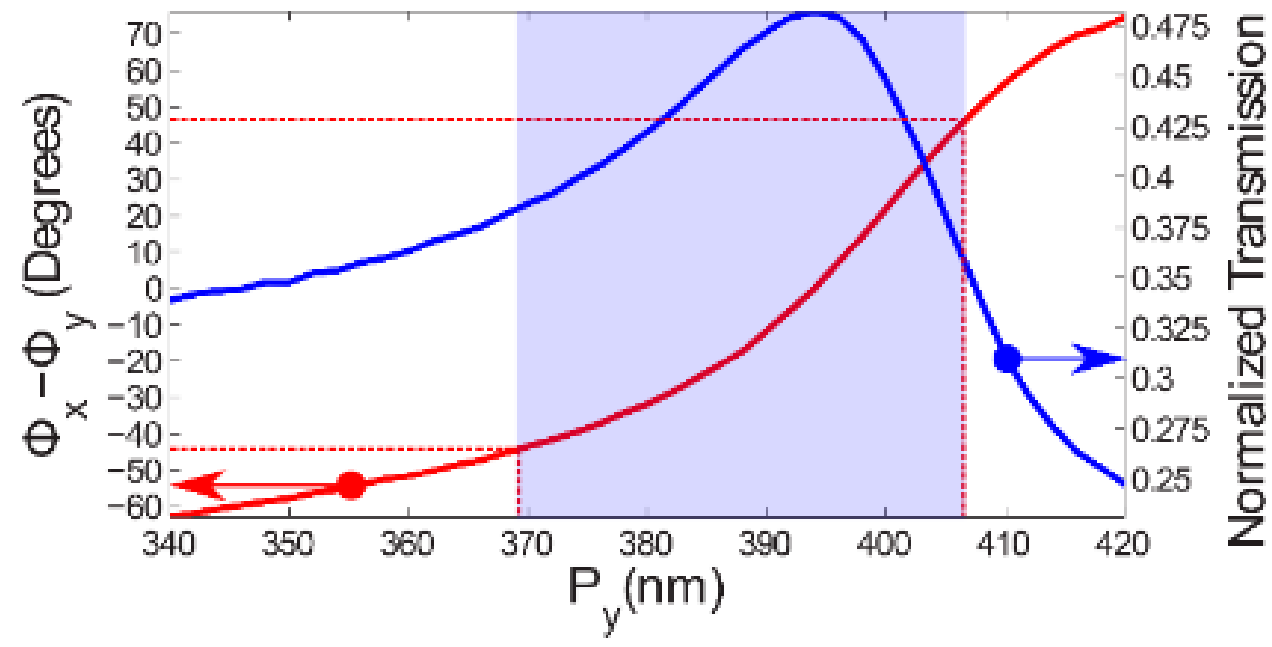

Figure 7: Relative phase differences between the *x* and *y* components of the transmitted electric field and Absolute transmission (normalized to the intensity over a unit cell) as a function of $P_y$ for $P_x = 394$ nm.

To satisfy condition **(B)**, a parametric sweep over the incident polarization was carried out and the state of polarization of the transmitted light was calculated using:

$$S_0 = |E_{tx}|^2 + |E_{ty}|^2 \qquad (17)$$

$$S_1 = |E_{tx}|^2 - |E_{ty}|^2 / S_0 \qquad (18)$$

$$S_2 = 2\,\mathrm{Re}\left[E_{tx}E_{ty}^*\right] / S_0 \qquad (19)$$

$$S_3 = 2\,\mathrm{Im}\left[E_{tx}E_{ty}^*\right] / S_0 \qquad (20)$$

where $E_{tx}$ and $E_{ty}$ are the transmitted *x* and *y* components of the electric field respectively obtained from simulations. The optimum incident polarization at $\lambda_0 = 700$ nm, was found to be $\alpha = 47°$, see Figure 8(a). To explain the optimal values for $\alpha \neq 45°$, consider the SPP fields propagating away from a single hole adhering to the complex phasor [25]:

$$E^{SPP} \propto \left(\hat{z} - \frac{i}{K_{SPP}}\rho\right) H_1^{(1)}(k_{SPP}\rho)cos(\varphi) \qquad (21)$$

where the strength of the SPPs is governed by $\hat{n}\cdot\mathbf{k}_{SPP}$, due to the dependency on $\cos(\varphi)$ [26], with $\hat{n}$ being the normal vector from the cavity to an observation point on the surrounding surface, see figure 29 in [21]. When considering a biperiodic array of holes, there is an optimum incident polarization angle where the SPPs are launched with equal amplitudes in two orthogonal directions, that is:

$$H_1^{(1)}\left(k_{SPP}P_x\right)\cos(\alpha) - H_1^{(1)}\left(k_{SPP}P_y\right)\sin(\alpha) = 0 \qquad (22)$$

The optimum incident polarization angle calculated from equation (22) was found to be $\alpha = 46.5°$ from the *x*-axis of the array and in agreement with that obtained numerically, see Figure 8(a). Same 3D numerical model described above was used to examine the optical response of the device when illuminated from the air and measured from the glass side. Absolute far-field transmission through the device, $P_t/P_0$, was calculated for $\alpha = \{0°, 47° \text{ and } 90°\}$, where $P_t$ and $P_0$ are the transmitted power, through the device and through the glass substrate in the absence of the device respectively, compare Figure 8(b) to Figure 6(b). With $\alpha = 47°$, transmitted Stokes parameters were calculated, compare Figure 8(c) to Figure 6(c). Numerically obtained DOP remained precisely unity for all wavelengths as expected. Slight red-shift in the $(1,1)_{glass}$ mode, for the change in incident polarization $\alpha = 0° \rightarrow 90°$, is somewhat unexpected. No matter the incident polarization, the $(1,1)_{glass}$ lattice mode is degenerate and common for all incident polarization. So, I would ascribe the change in momentum to the actual $(1,1)_{glass}$ lattice mode (that is invariant) being superposed with the quasi-$(1,1)_{glass}$ mode that varies with polarization. With the array being illuminated from the air side, one may argue that the strong $(1,1)_{glass}$ mode and its red-shift is due to the superposition of $(1,1)_{glass}$ and $(1,0)_{air}$ modes. However, I have demonstrated the optical response of an array of holes with $n_h = 1$, supported on a glass substrate, is dominated by glass modes regardless, please see section 6.3 of my thesis[21].

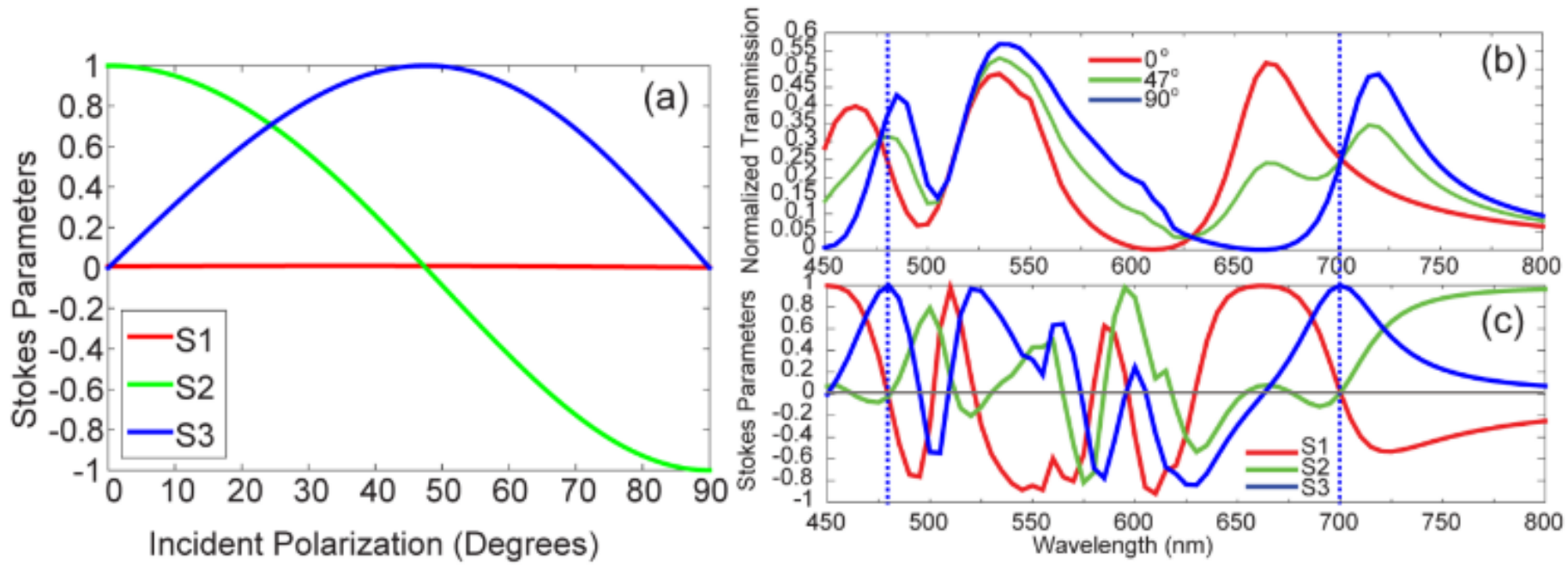

Figure 8: Spectra of the hole array having periodicities $P_x = 368$ and $P_y = 407$ nm. (a) Stokes parameters vs. the incident polarization. (b) Absolute transmission vs. the wavelength. Line-in-red relates to $P_x$ modes excited when α = 0°. Line-in-blue relates to $P_y$ modes excited when α = 90°. Line-in-green relates to both $P_x$ and $P_y$ modes excited when α = 47°. (c) Stokes parameters vs. the wavelength for α = 47°.

### *3.1 Significant Observations and Applications*

#### 3.1.1 Extraordinary Optical Transmission: (The Mechanism)

The two somewhat peculiar observations are: (1.a): *The formation of surface charges at the glass/silver interface when the device was illuminated from the air side where no SPP mode is supported at air/silver interface for the resonant wavelength $\lambda_0$ = 700 nm*. This was also highlighted in my theses, see section 6.3 [21]. (1.b): *The state of polarization of the transmitted light with* $S_3 = 1$ *and yet* $S_1 = S_2 = 0$ *when the incident light is linearly polarized at α = 47°*. From (1.a) and (1.b) one can develop a clear picture on the origin of Extraordinary Optical Transmission (EOT) [27] that is to say: *no light was transmitted through the hole directly*. Consequently, the transmission of power through the device must have pertained to the following steps: (i) excitations of LSPs inside the holes by the incident light, (ii) launching of the SPPs by the LSPs, hence LSP-SPP coupling and (iii) partial scattering of the LSP/SPPs in the form of free propagating EM waves. This is true at least in the case of a plasmonic hole array modelled here, when the dimensions of the hole were not optimized for the target wavelength. In other cases/devices, there may also be a direct transmission through the hole that superpose what I just described.

#### 3.1.2 Rotating Surface Charge Densities and Dipole Moments

What happens to the surface charges when both conditions **(A)** and **(B)** are satisfied? Figure 9(a) shows the entire unit cell and the transmitted electric field showing CPL state of polarization when the device was normally illuminated from the glass side with $\alpha = 47°$ and $\lambda_0 = 700$ nm. Figure 9(b)-(c) depicts the respective surface charge densities at silver/air and silver/glass interfaces. An extended model consisting of nine unit-cells in a 3-by-3 array formation was also simulated. Figure 9(d) represents the top view of the simulated surface charge densities at the silver/glass interface of the 3x3 model, when the device was normally illuminated from the air side with $\alpha = 47°$ at design wavelength $\lambda_0 = 700$ nm and at $t = \{0, T/8, T/4, 3T/8\}$, where $T$, is the period of the optical wave. See Appendix A, for method of calculating the surface charge density.

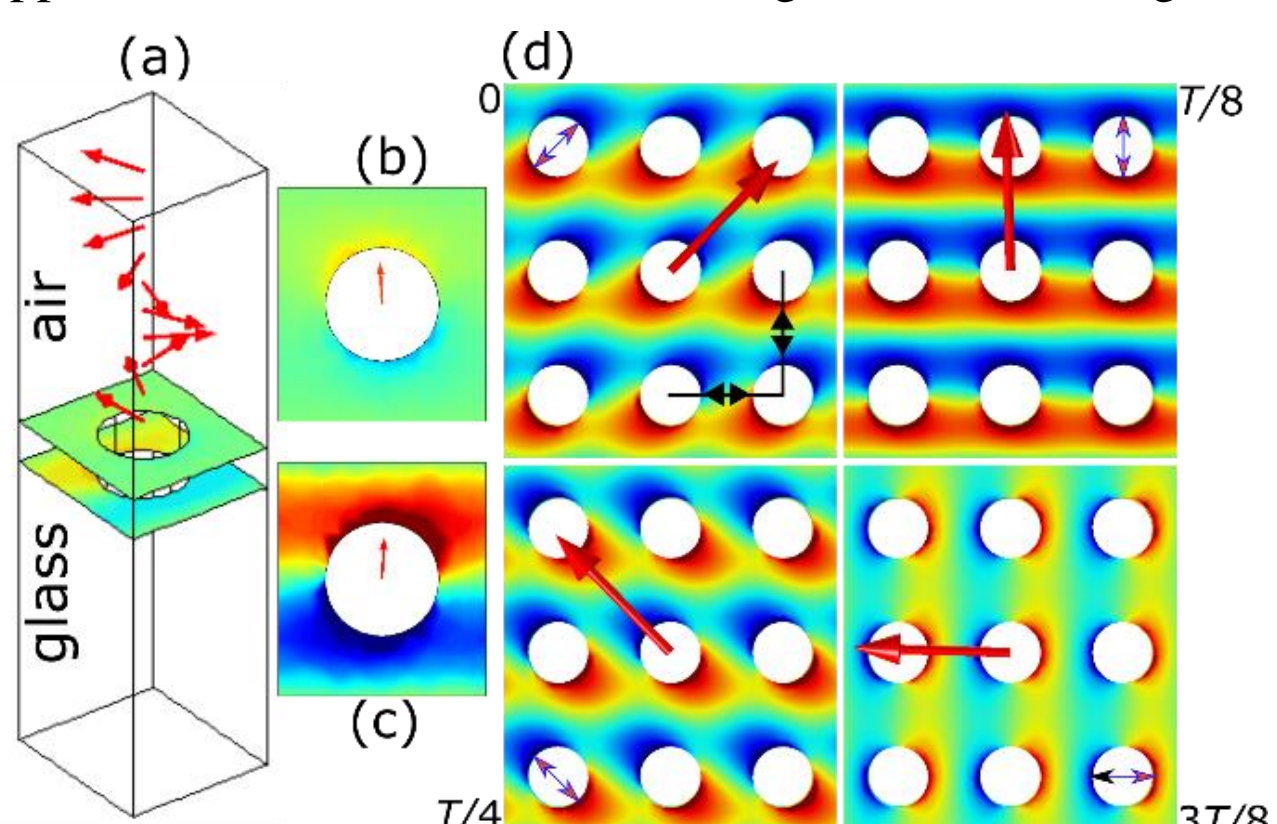

Figure 9: Surface charge density (red = 1, blue = -1 and green = 0) and the transmitted electric field vector represented by red arrows, calculated (a) within the entire of the unit cell (b) at air/silver interface of the unit cell, (c) at glass/silver interface the unit cell (d) at glass/silver interface of a 3x3 array, at t = {0, $T/8$, $T/4$, $3T/8$}, where $T$ is the period. Double arrow-heads show the orientation of the virtual dipole moments associated with the charge imparity formed on the rims of the holes.

The consequence of rotating surface charge density on the rims of the holes are (2.a): *the formation of rotating dipole moments,* $\mu_e$*, governed by equation* (2) *at the lattice points* and (2.b) *dipolar formations between a hole and its eight nearest neighbors that are discrete in time, occurring every T/8*. The latter being more peculiar than the former given that it occurs at $\lambda_0$ = 700 nm where the only supported mode is supposed to be $(1,0)_{glass}$, a clear breakdown of equations (7) and (8).

3.1.3 Three Level Quantum System

One may consider a biperiodic plasmonic hole array as a quantum interferometer with interesting properties. The most significant observation in that regard is (3.a): *the DOP = 1 for the transmitted light at all wavelengths*. If one considers the Stokes parameters as a probability amplitude of a quantum system, the fact that DOP remains unity for all wavelengths becomes significant. Owing to their strong correlation via equations (17)-(20) (or equations (9) & (11)-(14) in the case of analytical solutions), a quantum system that satisfy probability sum $S_1^2 + S_2^2 + S_3^2 = 1$, doesn't suffer from *depolarization* (that is a form of quantum decoherence). Therefore, above model with its two orthogonal modes is well suited to describe a *three-level quantum system for two (or more) entangled photons*. Consequently, for a given incident polarization and wavelength, ($\alpha$, $\lambda_0$), one may define the three-level mixed state of polarizations for multi photons, as:

$$\left|\psi_{(\alpha,\lambda)}\right\rangle\Big|_{x,y=0} = S_{1(\alpha,\lambda)}\left|+\right\rangle + S_{2(\alpha,\lambda)}\left|\times\right\rangle + S_{3(\alpha,\lambda)}\left|\bigcirc\right\rangle \qquad (23)$$

where $|+\rangle, |\times\rangle \,\&\, |\bigcirc\rangle$ represent the *single* level pure states of polarization with $S_1, S_2, S_3$ being the probability amplitudes that range from -1 to 1. Each pure state encapsulates the $x$ and $y$-polarized photons with $|+\rangle \equiv |x\rangle$, $-|+\rangle \equiv |y\rangle$, $\pm|\times\rangle \equiv \frac{1}{\sqrt{2}}\left(|x\rangle \pm |y\rangle\right)$ and $\pm|\bigcirc\rangle \equiv \frac{1}{\sqrt{2}}\left(|x\rangle \pm i|y\rangle\right)$, compare the last two definitions to equations (3) -(4). The negative probability amplitudes do not violate the superposition principle in (23). With that in mind, Figure 8(c) maps all possible states in Hilbert space, for this particular hole array, that is for the given choice of incident polarization, material and dimension as stated in the previous section. Now, what formalism/algorithm must be implemented based on equation (23) and how this could play a role in quantum computation/information/cryptography deserves further study but is beyond the scope of this report. Notably, every mixed state $\left|\psi_{(\alpha,\lambda)}\right\rangle$ indexed by a particular incident polarization and wavelength, ($\alpha_i$, $\lambda_j$), has its own probability amplitude, governed by the normalized transmission $I = P_t/P_0$. Consequently, when the device is excited with an incandescent incident unpolarized light, the state of the system as a whole must be denoted by:

$$\psi_{system} = \sum_{i,j} c_{(\alpha_i,\lambda_j)} \left|\psi_{(\alpha_i,\lambda_j)}\right\rangle \qquad (24)$$

where the normalized portability amplitudes are given by:

$$c\left(\alpha_i,\lambda_j\right) = I\left(\alpha_i,\lambda_j\right) \Big/ \int_{\alpha_i}^{\alpha_f}\int_{\lambda_i}^{\lambda_f} I\left(\alpha,\lambda\right) d\lambda\, d\alpha \qquad (25)$$

Equation (24) implies that the state of the system as a whole is further, the result of superposition of individual states $\left|\psi_{(\alpha_i,\lambda_j)}\right\rangle$. But do they form an orthonormal basis on their own? I will address this question in the experimental section. However, for the particular incident polarization $\alpha = 47°$ and the wavelength range that produced the spectrum in Figure 8(b)-(line in green), equation (25) maybe reduced to $c\left(47°, 700nm\right) = I\left(47°, 700nm\right) \Big/ \int_{450nm}^{800nm} I\left(\lambda\right) d\lambda$. On the other hand, one may even add another parameter to equation (25) by considering the angle of incidence. So, there is room to play.

3.1.4 Applications in Gravitational Decoherence

Recently I came across two articles on "Gravitational Decoherence".

Anastopoulos and Hu [28] defining the intrinsic or fundamental decoherence as, and I quote: "*Intrinsic or fundamental decoherence refers to some intrinsic or fundamental conditions or processes which engender decoherence in varying degrees but universal to all quantum systems. This could come from (or could account for) the uncertainty relation, some fundamental imprecision in the measuring devices (starting with clocks and rulers), in the dynamics , or in treating time as a statistical variable*" [28].

Bachlechner [29] wrote: "*The decoherence effect can be modeled as a quantum Zeno effect in which the wave function of the tunneling field "collapses" to a classical configuration each time the background leaks information to the environment about whether a bubble exists or not*."[29].

DOP = 1 in the model I report here, infers a quantum system that is deterministic with no *uncertainty relation*. This is true at least in theory given that the numerical model of the array represents an ideal system that interacts only with light when there is no *imprecision in the measuring devices* or any other external factor causing decoherence. I am not an expert in gravitation but if I understood the two reports correctly, I would propose that the device I reported here, may have an application in detecting gravitational decoherence by monitoring and anticipating the collapse of the wave function that would result in DOP fluctuations.

Furthermore, considering two identical biperiodic hole arrays {*array*1, *array*2}, acting as a transmitter and a receiver, when the first device is excited by a linearly polarized light, and the resulting transmitted CPL be an input to the second device, the transmitted light through the second device would be a linearly polarized light with the original incident angle of polarization, as long as the liked-sides of the arrays, (i.e. substrate-to-substrate or superstrate-to-superstrate), are paired in this interaction. In other words:

$$\left(A|+\rangle+B|\times\rangle\right)\overset{input1}{\rightarrow}\left[array1\right]\overset{output1}{\rightarrow}|\bigcirc\rangle\overset{input2}{\rightarrow}\left[array2\right]\overset{output2}{\rightarrow}\left(A'|+\rangle+B'|\times\rangle\right) \quad (26)$$

where $A' = A \,\&\, B' = B$ , which was confirmed numerically for normally incident light. This is intuitive since the outgoing $\pm|\bigcirc\rangle$ polarized light propagating with $+\mathbf{k}_0$ away from *array*1, impinges on *array*2 with $-\mathbf{k}_0$ hence analyzed as having opposite spin, $\mp|\bigcirc\rangle$ , and in the process the amplitude ratio and phase are restored to that of the original incident light. In effect, *array*2 must be considered as the time-reversal operator of *array*1. Deployment of two identical arrays positioned far apart, coupled via $|\bigcirc\rangle$ , must be looked upon as two arrays of quantum interferometers that are coupled. In this case, any gravitational decoherence impacting the optical responses of either or both devices, would result in a final state other than the initial state, where $A' \neq A$ and/or $B' \neq B$ . The motivation of implementing two coupled quantum optical device is to increase the sensitivity of the system as a whole with respect to gravitation decoherence.

3.1.5 Applications in Detecting Gravitational Waves

Fundamental physics governing Michelson interferometers is that of the interference whereby the passage of the two monochromatic beams of the same wavelength through equal/unequal optical paths and their superposition afterwards would create either constructive or destructive interference at the point of detection. Circularly polarized light in conjunction with biperiodic arrays may play an important role in detecting gravitational waves as well as revealing more about their nature. The question is by how much the two conditions **(A)** and **(B)**, as stated in the introduction, are violated by gravitational waves. Furthermore, considering the second part of equation (26), i.e. $|\bigcirc\rangle\overset{input2}{\rightarrow}\left[array2\right]\overset{output2}{\rightarrow}\left(A'|+\rangle+B'|\times\rangle\right)$, it is apparent that the role of the array is that of a polarization converter, that is to convert an incoming CPL into an outgoing linearly polarized light (LPL). It is, therefore, possible to design an interferometer based on polarized light that would isolate the two orthogonal polarization states, hence study the impact of gravitational waves on each type of polarization. It is also possible to further analyze the impact of the gravitational effect on electromagnetic as well the electronic activities [30-35], governing such interferometer.

An alteration to Michelson interferometer based on polarized light is depicted in Figure 10. The black box labelled as Source/Splitter/Polarizer/Combiner, generates a coherent and monochromatic light which is then split into horizontally/vertically polarized lights (denoted by $\alpha = 0°$ and $\alpha = 90°$ in the figure), with each type polarization being routed to the relevant arm of the interferometer. Armlengths with $L_1 \neq L_2$ are set in such ways to bring about the 90° phase difference between the two beams, upon their reflection back to the black box, i.e. after having travelled $2L_1$ and $2L_2$ . The two beams are then superposed to produce a single beam that is circularly polarized. CPL is then forwarded to a detector consisting of a plasmonic biperiodic array (i.e. an array of quantum interferometers) that converts CPL to LPL with a well-defined polarization axis. An analyzer with its polarization axis adjusted to the incoming LPL, serves as filter allowing/blocking the passage of photons to the counter. Any changes to the beam along the $L_1$ or $L_2$ would result in distortion of the resultant CPL in the form of induced ellipticity which in turn would result in the change in the polarization axis of photon that fall on the analyzer, hence change in photon count. The amplitude of the interference also plays the same role as Michelson interferometers which also impact the number of photon detected. In the figure I have considered two distinct scenarios, when the gravitational is either normal to $L_1$, or to $L_2$. Apart from a new interferometer design, the aim is to study the impact of gravitational waves on polarized of light.

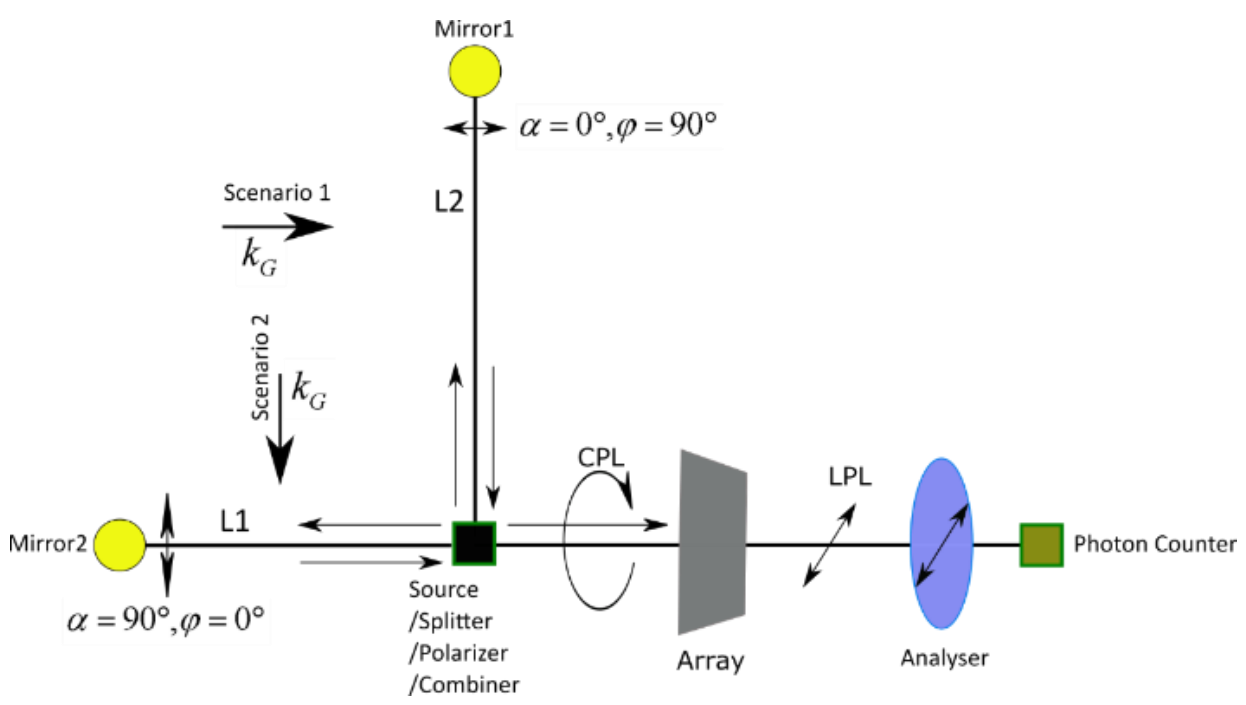

Figure 10: Schematics of an interferometer based on polarized light with a plasmonic biperiodic array and an analyser and a photon counter as an integrated detector.

*3.2 Experimental Demonstration*

The device was first fabricated with a silver film on a glass substrate and was reported in [21,24]. Results are depicted in Figure 11.

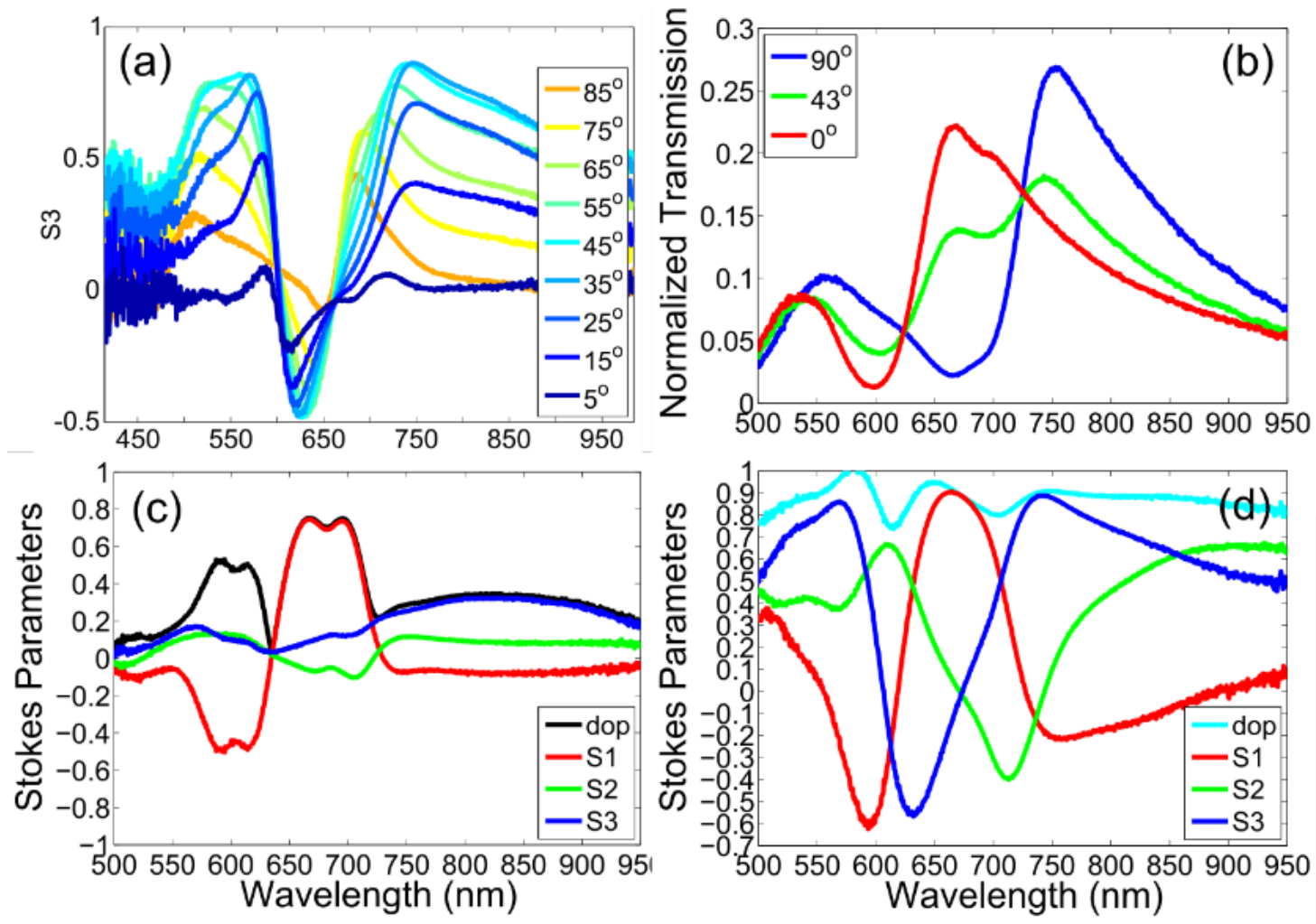

Figure 11: Experimental data: (a) Transmitted $S_3$ spectra for incident polarizations $5° \leq \alpha \leq 85°$. (b) Absolute transmission vs. the wavelength. Line-in-red relates to $P_x$ modes excited when $\alpha = 0°$. Line-in-blue relates to $P_y$ modes excited when $\alpha = 90°$. Line-in-green relates to both $P_x$ and $P_y$ modes excited when $\alpha = 43°$. (c) Stokes parameters vs. the wavelength for incident polarization 43°. (d) Stokes parameters vs. the wavelength for un-polarized normally incident light.

Firstly, I must provide an explanation on the discrepancy between the incident polarization obtained numerically vs that obtained experimentally, that is $\alpha = 47°$ vs $\alpha = 43°$ respectively. Examining the numerical results in Figure 8(b), notably the peak intensity of (1,0) mode associated with $P_x$ is slightly stronger than that of (0,1) related to $P_y$ with the ratio of their peak intensities being $I_{(0,1)}/I_{(1,0)} \approx 0.9$. Naturally, in satisfying the condition **(B),** the incident polarization must deviate from the *analytically* obtained $\alpha = 46.5°$, in favoring the weaker mode, (0,1), by a slight increase in α. In other words, the extra 0.5° adjustment from $\alpha = 46.5°$ to $\alpha = 47°$, towards the *y*-axis compensated for the weaker (0,1) mode to satisfy condition **(B)**. In the case of the experimentally obtained spectra, Figure 11(b), the ratio is $I_{(0,1)}/I_{(1,0)} \approx 1.23$, consequently the deviation from the analytical α = 46.5° must be toward the *x*-axis, hence a total adjustment of -3.5° resulting in α = 43°. I can also refer the reader to my numerical results in Figure 1(a) where the disparity between the two RCSs were so strong that I had to set α = 32° to satisfy condition **(B)**. Calculating the *adjustment* to α analytically, depends on few factors including the spectral line-shapes (be it obtained analytically or experimentally) and somewhat tedious. But the main point here is that by adjusting *α* it is possible to obtain the maximum degree of circularly polarized light and this is clearly shown in Figure 11(a).

Earlier simulations showed improvement in transmission through a hole array when it is set in a homogeneous environment with the substrate, superstrate and the hole having the same refractive indices (i.e. $n_1 = n_2 = n_3$), see section 6.2 of my thesis. Same device was converted into a freestanding array after being etched with hydrofluoric acid. This would allow for the integration of the array into a homogeneous environment. Note the silver film thickness of 71 nm as measured by SEM, see Figure 12. Although optical measurements were taken prior to the cross-sectional images, film thickness of more than 50 nm was a clear indication that leakage of incident field through the film was either negligible, or none at all, during my optical measurements.

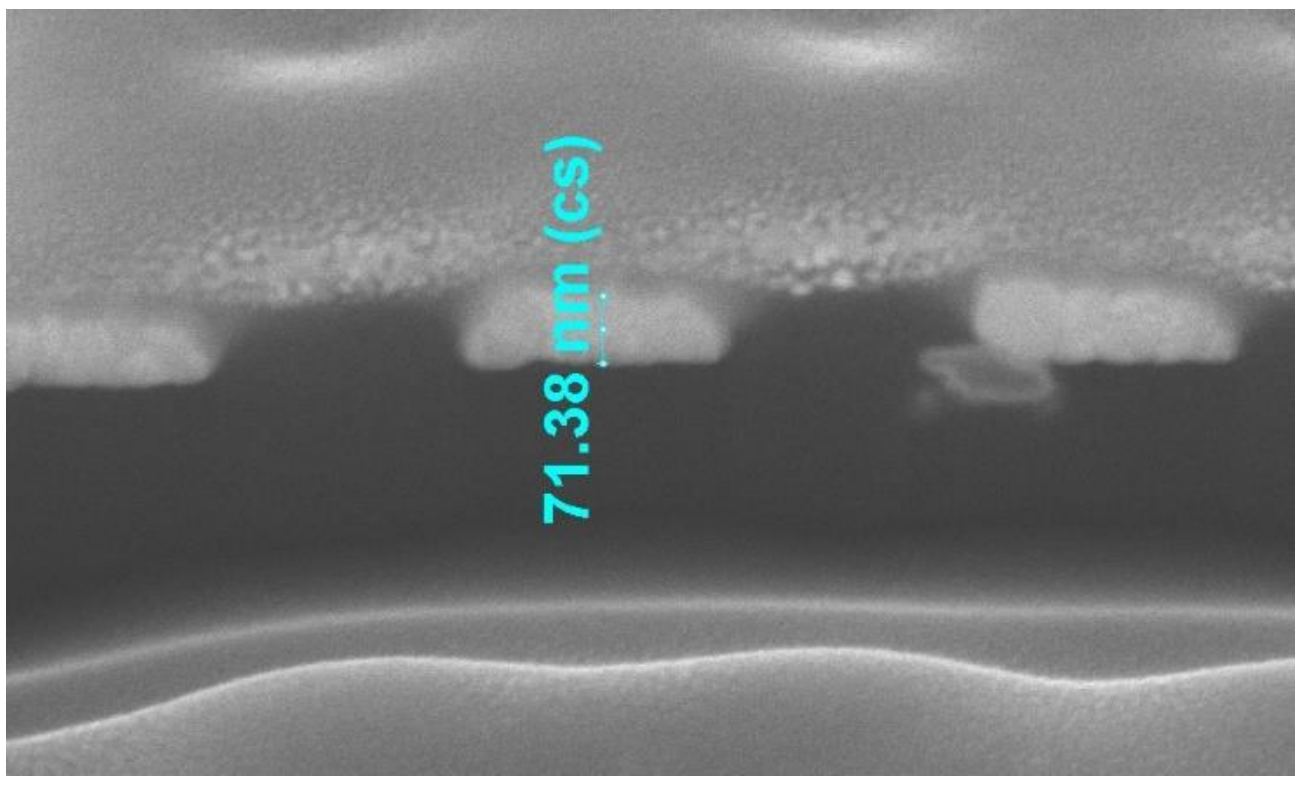

Figure 12: Cross-sectional SEM image of the biperiodic array of holes as reported in [36] freestanding F2:2 array in a homogenous dielectric environment (immersion oil). F2:3(d) Close-up of (c).

As for transmitted CPL, previous experiment had resulted in $S_3 < 1$, that contradicted my numerical results. So, I implemented the technique suggested by Kihara [37] that catered for phase errors associated with the optical elements, that confirmed $S_3 = 1$ is indeed possible, see Figure 13. With the new device, criteria **(I)- (II)** were satisfied in full. I have elaborated on the fabrication method, measurement techniques, numerical/experimental results and its application as a refractive index sensor in details previously [36].

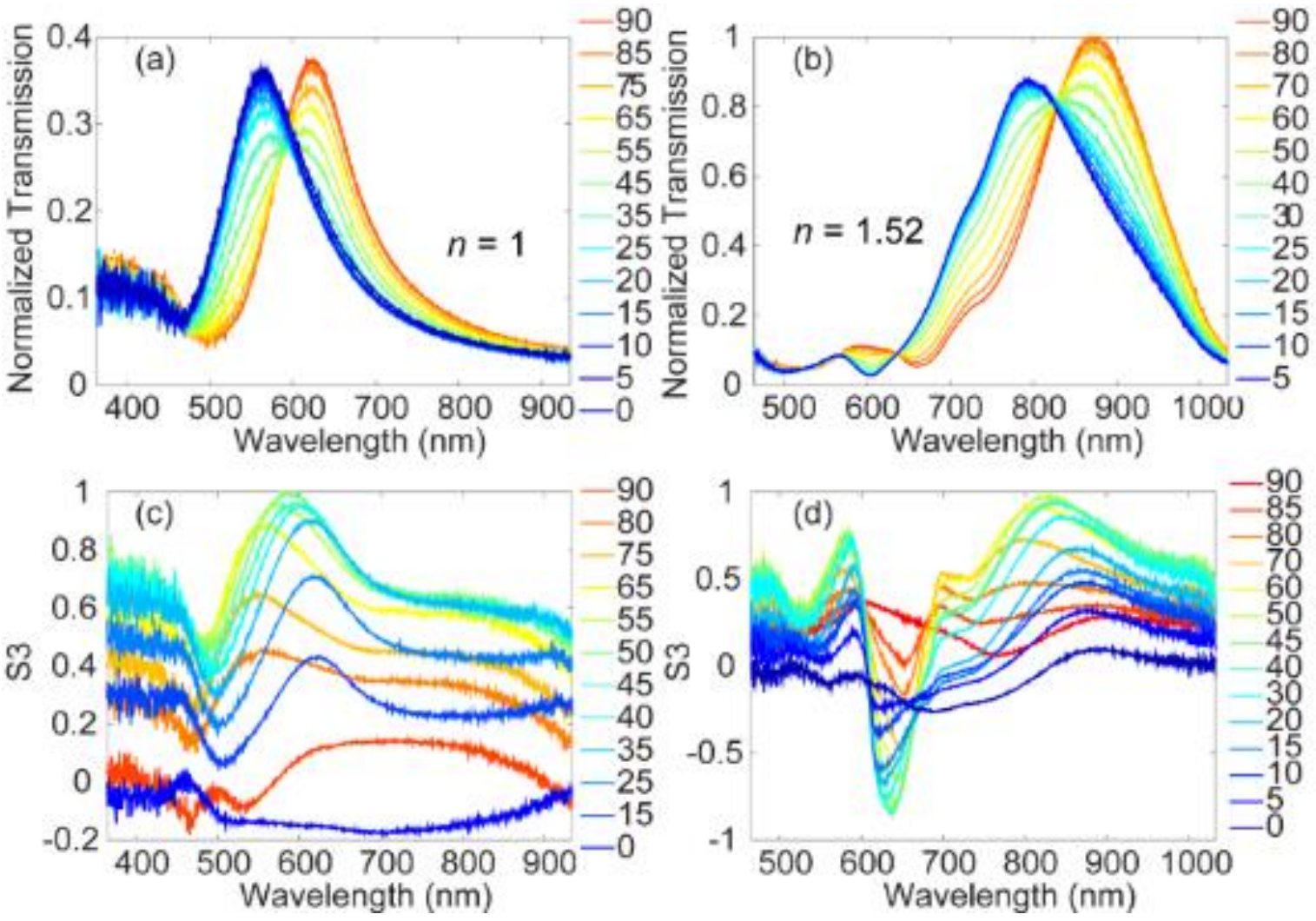

Figure 13: Experimentally obtained results for incident polarization $0° \leq \alpha \leq 90°$. Absolute transmission vs. the wavelength for (a) $n_1 = n_2 = n_3 = 1$ and (b) $n_1 = n_2 = n_3 = 1.5$. $S_3$ vs. the wavelength for (c) $n_1 = n_2 = n_3 = 1$ and (d) $n_1 = n_2 = n_3 = 1.52$.

Obviously, immersing the array in a homogeneous medium have improved its quantum efficiency, sufficient for emission when considering part of the design represented by the left side of equation (26). Any further concerns on plasmonic losses and quantum efficiency when detecting such photons using the same array as proposed by the right side of equation (26), may be addressed by implementing the array within a homogeneous gain material, rather than one whit $n = 1.5$. Given that photons produced during the stimulated emission, follow the phase and polarization of the stimulating photon, I am confident that the use of gain material does not violate the governing principles on polarization effects in this report. Hypothetically, it is possible to adjust the intensity of the incident beam for two entangled photons be emitted at a time, eliminating the need for any integrated single photon emitters. Note that my definition of two photons being in an *entangled state*, refers to a single quantum system, be it vacuum, consisting of two photons with their coherent time and length extended to the far-field if not infinity, yet representing a polarization state that is only possible while the constituting photons remain part of that system, interacting with one another via their individual orthogonal polarizations.

3.2.1 Encapsulation of Three Level Quantum State into a Continuous Orthonormal Set

Notably, analytical and numerical results were carried out with a single wavelength at the time, whereas the experimental results were obtained by exciting the array using a white light. Therefore, it was important to determine whether multiwavelength excitation of the array had any impact on the spectra. In the experiment I report here, the array was illuminated from both sides using two independent light sources. From the glass/silver side the array was illuminated with a Fianium Supercontinuum light source filtered at $525 < \lambda_0 < 560$ nm. The filter also allowed partial transmission at $435 < \lambda_0 < 450$ nm. These coincided with the $(1,1)_{\text{glass}}$ and $(0,2)_{\text{glass}}$ modes along the $x$-direction. Number of photons detected from the Fianium Supercontinuum light source passing through the filter vs the wavelength in the absence of the device is depicted in Figure 14(a) and inset. Spectra in red and green are produced with $q = 400$ and $q = 420$ respectively. Here, $q$, is the quality factor that determines the output power and the spectral line profile of the Fianium's emission, with $q = 0$ meaning no emission. With the device in place, in order to excite all modes associated with $P_x$, the array was first illuminated normally from the air/silver side with a halogen light polarized at $\alpha = 0°$. The Fianium power was then changed from $q = 0$ to $q = 400$ to $q = 420$. Figure 14(b) shows the transmission spectra of the polarized halogen light through the device superposed by reflection spectra of the filtered Fianium light source from the device as detected by the spectrometer. Figure 14(c)-(d) are the selected regions from Figure 14(b). This exercise is not about percent reflection/transmission from/through the device. The main motivation behind this work was to detect any noticeable change in amplitude or phase, in the $(1,0)_{\text{glass}}$ when $(1,1)_{\text{glass}}$ or $(2,0)_{\text{glass}}$ modes are perturbed. Jitters observed in the $(1,0)_{\text{glass}}$ , Figure 14(d), are present for all $q$ values, hence attributed to the thermal instability of SPPs and perhaps the optical instruments and light sources. Considering the photon counts by which the $(1,1)_{\text{glass}}$ is perturbed vs the perturbation observed in $(1,0)_{\text{glass}}$, it is safe to conclude that although multiple SPP-Bloch modes can coexist, coupling between them (if any) is insignificant. This has an important ramification, that is $\left\langle \psi_{(\alpha,\lambda_i)} \middle| \psi_{(\alpha,\lambda_j)} \right\rangle = \delta_{ij}$ . Depending on how one considers the wavelength divisions, states $c_{(\alpha,\lambda_i)} \left| \psi_{(\alpha,\lambda_i)} \right\rangle$, can now form a *continuous orthonormal set*, with each $\left| \psi_{(\alpha,\lambda_i)} \right\rangle$ perceived as a pure quantum state on its own.

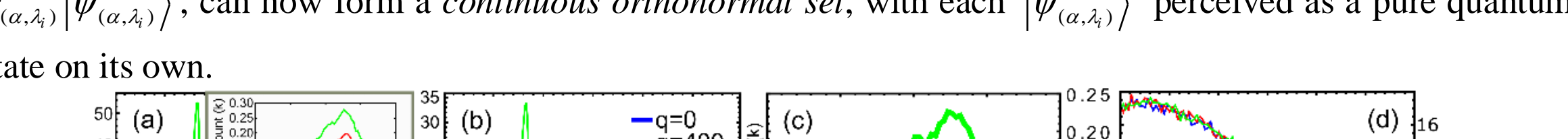

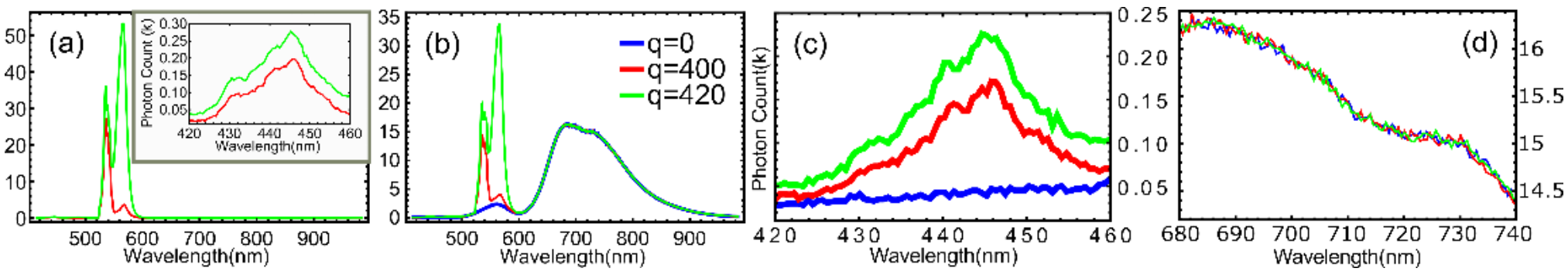

Figure 14: (a) Photon count detected from the Fianium Supercontinuum light source passing through the filter vs the wavelength in the absence of the device. (b) Transmission spectra of the polarized halogen light through the device superposed with the reflection spectra of the filtered Fianium light source from the device as detected by the spectrometer. Spectra in blue, red green are produced with q = 0, q = 400 and q = 420 respectively. (c) and (d) selected regions from (b).

## 3. An Argument on Photon Spin

The most important aspect of the experimental setup in this report was the use of Andor Shamrock 303i-A spectrometer (with Andor iDus DU920P-BR-DD CCD) which I believe is a highly accurate instrument in counting the number of photons per frequency/wavelength. And notably, although the spectra were normalized, they are indicative of photon counts. This is important as I will shortly address the dilemma I was facing with respect to quantum presentation of the spin of a photon.

Although, in my opinion, it is acceptable to use $\pm\hbar$ as a *notion* to refer to $\pm|\bigcirc\rangle$ polarization state of *light*, as I have done in section 2.3 of my master thesis [9], I do not support the idea of a *single* photon having a $+\hbar$ or $-\hbar$ spin, for example as asserted in section "8.1.4 Angular Momentum and Photon Picture" in [38]. In classical optics, it is a well-known fact that the a linearly polarized light may be constructed from the superposition of two opposite circularly polarized light. In providing an explanation for linearly polarized photons, however, Hecht applies a similar principle whereby each photon in a linearly polarized light must exist in either $+\hbar$ or $-\hbar$ state with equal probabilities, as nothing is known about the state of a photon before measurements, by stating, and I quote: "*We cannot say that the beam is actually made up of precisely equal amounts of well-defined right- and left-handed photons; the photons are all identical. Rather, each individual photon exists in either spin state with equal likelihood. If we measure the angular momentum of the constituent photons, $-\hbar$ would result as often as $+\hbar$ ...* "[38]. To my opinion, while the classical picture of linearly polarized light is valid, the latter is flawed, as it implies that somehow optical elements dictates the spin of a photon upon measurements. Note, a 50/50 percent chance of a photon being in $+\hbar$ or $-\hbar$ state, best fits the description of an unpolarized light prior to polarization but not a linearly polarized photon before being measured or analyzed. Hecht hypothesize further on the origin of circularity/ellipticity of light in terms of photons by stating: *"In contrast, if each photon does not occupy both spin states with same probability…The result en masse is elliptically polarized light, that is, a superposition of unequal amounts of R- and L-light bearing a particular phase relationship bearing a particular phase relationship"* [38]. Implying that the mixed state of polarization of light, that is the *ellipticity* which is the superposition of circularly and linearly polarized lights, is purely due to unequal number of photons being in $+\hbar$ and $-\hbar$ spin, hence in contrast to what I have proposed in equation (23) based on the $x$- and $y$-polarized photons.

Such disparities between quantum and classical picture of photon and light has created confusion ever since the introduction photon. The quantum picture of photon spin is not unique to Hecht [38], and I intend no criticism of any author, as I believe Hecht and others were merely inferring and reporting on the scientific consensus of their time. Referencing a very popular text in optics has historical reasons as the statements quoted from Hecht [38], had a direct impact on development of ideas in this report, for if there was no ambiguity, there would not be a quest to resolve it. But I hope a clear picture of the polarization of photons would emerge as the result of this report.

I assert that the optical response of the device to the incident light polarized at $\alpha = 0°$ and $\alpha = 90°$, as seen in Figure 11(b) and Figure 13(a)-(b), corresponds to all $x$-polarized and $y$-polarized photons with no spin. The line in blue as seen in Figure 13(b), for example, represents the population density per wavelength for all the $x$-polarized photons emitted by the device and detected by the instrument, when the incident light was polarized at $\alpha = 0°$, whereas the line in red corresponds to all the $y$-polarized photons when the device was excited with light polarized at $\alpha = 90°$. However, the two distinct linearly polarized spectral response of the array says nothing conclusive about the nature of photons or their spins, as one may still argue that a linearly polarized incident or transmitted lights are due to equal number of photons in $+\hbar$ or $-\hbar$ spin states with a particular phase difference as stated by Hecht [38].

But that is not the case given the experimentally obtained Stokes parameters vs. the wavelength when I used normally incident unpolarized light to excited the array, see Figure 11(c). When resonant transmission mode of the array in the $x$-direction overlaps with the suppressed transmission mode in the $y$-direction, and vice versa, randomly polarized photons in an incident unpolarized light are filtered in such a way that solely photons that are polarized along the direction of resonant modes make contribution to the transmission. Transmitted $S_1$ = DOP = 0.8, as seen in Figure 11(c), in the regions of $660 \leqslant \lambda_0 \leqslant 700$ nm is due to the resonant mode of the array in the $x$-direction at $\lambda_0 = 625$ nm, excited in response to all (or the majority of) the incident $x$-polarized photons that existed within the unpolarized incident light, while the suppressed mode at the same wavelength blocks most $y$-polarized photons. Similarly, $S_1$ = -DOP = -0.6 in the range of $580 \leqslant \lambda_0 \leqslant 620$ nm, corresponds to all $y$-polarized photons exiting the array modes in the $y$-direction and so forth. It is becoming clear that an unpolarized light is an ensemble of randomly oriented linearly polarized photons with no spin. However, to obtain a conclusive evidence on photon's polarization, Figure 11(a)-(d) must be analyzed in aggregation. If one still persists on the existence of spin in a single photon, (*by arguing that the presence of x-polarized or y-polarized light within the incident unpolarized light is merely due to the formation of photon ensembles with equal number of +ħ and -ħ spins and a certain phase relationship between the two that leads to a linearly polarized light*), then I would state, when considering an unpolarized light, I see no difference between the direction defined by the *x*-axis and a direction defriended by 43° from the *x*-axis. Meaning, there should have been a set of photons forming a similar ensemble resulting in a linearly polarized light along $\alpha = 43°$. And if that was the case, then I should have obtained $S_3 \approx 1$ at $\lambda_0 \approx 740$ nm in Figure 11(c), just as I did in Figure 11(d). But clearly, that was not the case. Consequently, we must either accept that a single photon has no spin, or invalidate every experiment based on vertically/horizontally polarized photons, such as those reported in chapter 14 of Quantum Optics by Fox [4].

Going back to Figure 13(b), notably the two curves intercept at $\lambda_0 = \{820, 640, 580\}$ nm. Clearly these are the wavelengths at which the device is capable of producing equal (or near-equal) number of both *x*- and *y*-polarized photons. This condition is satisfied for some incident polarization, $45° \leqslant \alpha \leqslant 50°$, and given the phase difference between the two kinds of photons, $S_3 = \{1, -0.9, 0.8\}$ are achieved respectively. I can then infer that the photonic requirements for producing a circularly polarized light, hence its resultant electric field rotating about $\mathbf{k}_0$, are just the same as those described in the introduction, i.e. the adherence to conditions **(A)** and **(B)**. This means that we need at least two photons having same energies, but one vertically and the other horizontally polarized, with a time difference of $\Delta t = \lambda_0/4c$ between their propagations, where "c" denotes the speed of light in vacuum. Assuming that such pair were generated by the same quantum system, this also implies that one photon must lead the other one by the time difference, $\lambda_0/4c$, hence being detected earlier. If it is possible to analyze the polarization of such a pair at a higher sampling frequency than $(\lambda_0/4c)^{-1}$, I would hypothesize that the detected electric field remains linear during the first $\lambda_0/4c$ seconds, before starting its rotation about $\mathbf{k}_0$.

And finally, with the new proposed picture of photons lacking spins, a question may arise concerning light-matter interactions and the mechanism behind the transfer of angular momentum from a photon to a charged particle such as electrons in orbits. From the experimental data presented in this report (with the incident light being either polarized or unpolarized), it is evident that it is up to the *matter* to sort out (or in the case of CPL "pair up") the incoming photons. In other words, it is the *matter* with all its intrinsic physical properties that operates on the incident light, just as I described when explaining equation (1) above. And if there are experimental evidences that truly a *single* photon has induced changes to an electron's orbital angular momentum in an atom, then I would imagine, (and nothing more at this stage), that the change has to do with the absorption of photon's energy, and in the process the electron with its spin and angular momentum, as well as the atom hosting the electron with its intrinsic properties, have resolved the absorbed energy into observable changes to the orbital angular momentum.

I have retained my raw data files showing the photon counts …etc. and have revisited them multiple times in search for new findings and I will continue to do so. Upon any new discovery in contrary to what I have stated here, I will make the required adjustments.

## 5. Conclusions

I have discussed the concept "resultant dipole moment" starting from a simple asymmetric cross-shaped nano-antenna to a simple biperiodic array of holes. A clear link between resultant dipole moment, LSPs, SPPs, and the transmitted Stokes parameters was established, hence the Dipole-LSP-SPP-Stokes coupling. I have shown how a simple analytical model based on superposition of waves between two holes can predict plasmonic lattice modes. In addition, the analytical model identifies quasi-modes which explains the anomalous shift in (1,1) modes. Stokes parameters obtained analytically are in accordance with those obtained numerically. The cyclic plasmonic wavevector and the subsequent cyclic plasmonic dipole moments in hole arrays are further the direct consequent of the superposition of two orthogonal surface waves when S3 = 1. Similarly, the spiral surface charge densities surrounding an asymmetric cross is due to the two orthogonal aperture modes giving rise to a rotating dipole moment that launches SPPs in directions that changes with respect to time. When examining the surface waves, a biperiodic array with two unequal pathlengths, functions as an array of coherent quantum interferometers. I have explained how the state of polarization of such array may be considered as states in a three-basis quantum system. Furthermore, I have shown that SPP-Bloch modes do not interact with one another. That may provide an opportunity for the realization of a quantum system with a continues orthonormal set, i.e. a quantum system with many-basis. I have also highlighted aspects of the device applicable to generation/detection of entangled photons, and how any two such array may be coupled to achieve that. I also hypothesized on a possible application in gravitational detection/decoherence, however, what was suggested was a mere possibility hoping to ignite interest in bridging plasmonics to other field of physics. And finally, I have provided an argument on the polarization of a single photon being linear, hence lacking any spin. This is an enforcement of the proposed "minimum two entangled photon needed to produce circularly polarized light".

**Funding:** This research received no external funding

**Conflicts of Interest:** The author declares no conflict of interest.

## Appendix A: Calculations of Surface Charge Densities from Converged Numerical Solutions

Given that the electric field, **E,** is the *parameter to solve for* in all relevant FEM simulations here, all other subsequent physical quantities, including surface charge densities, must be derived from the electric field after a solution has converged. Method proposed here was inspired when revisiting **Fig. 6-6** and **Fig. 9-5** as depicted in *Electromagnetic Fields and Waves* by Lorrain [10]. Relationship between the normal to the surface component of the electric field, $E_z$, and surface charges are depicted in Figure A 1(a). Total surface charge density is given by $\sigma_{total} = \sigma_f + \sigma_b$, where $\sigma_f$ is the free charge density inside the metal and $\sigma_b$ is the bound charge density in the dielectric. One may obtain $E_{z\text{-}solution}$ at an interface directly from a converged solution, however, I am not certain how third-party FEM software packages (a black box) calculates $E_z$ at a metal/dielectric interface. I suspect that they adhere to the superposition principle, hence adding the field components at a mesh point, which is perfectly valid when the only quantity of interest is the total field itself, i.e. $E_{z\text{-}solution}$. But when the total surface charge density is to be obtained, $E_z$ must account for charges of opposite signs on both sides of the metal/dielectric interface, therefore the fields must be segregated accordingly. Concerning the field contributions from two opposite point charges to a midpoint on an interface in between, see Figure A 1(b), it is clear that $E_z = E_{z1} - E_{z2}$. The "-" signs ensure summation of the two fields when the charges are of different signs and their cancellation when they are of the same sign, hence $\sigma_{total} = \epsilon_0 \left( E_{z1} - E_{z2} \right)$ as I have stated in my thesis arXiv:1912.10888 [physics.optics], [21]. This is not exactly a text book approach, but I believe it is more definitive than just obtaining $E_{z\text{-}solution}$ over an interface from a converged solution. Most finite element software packages have built-in to calculate the electric fields at either side of an interface separately. In summary, after a solution is converged, one must obtain $E_{z\text{-}solution}$ directly at the interface and then use the built-in functions (let's call them A() and B()) to isolate the fields corresponding to each side, e.g. $E_{z1} = \mathrm{A}(E_{z\text{-}solution})$ and $E_{z2} = \mathrm{B}(E_{z\text{-}solution})$, then apply $\sigma_{total} = \epsilon_0 \left( E_{z1} - E_{z2} \right)$ to obtain total surface charge density.

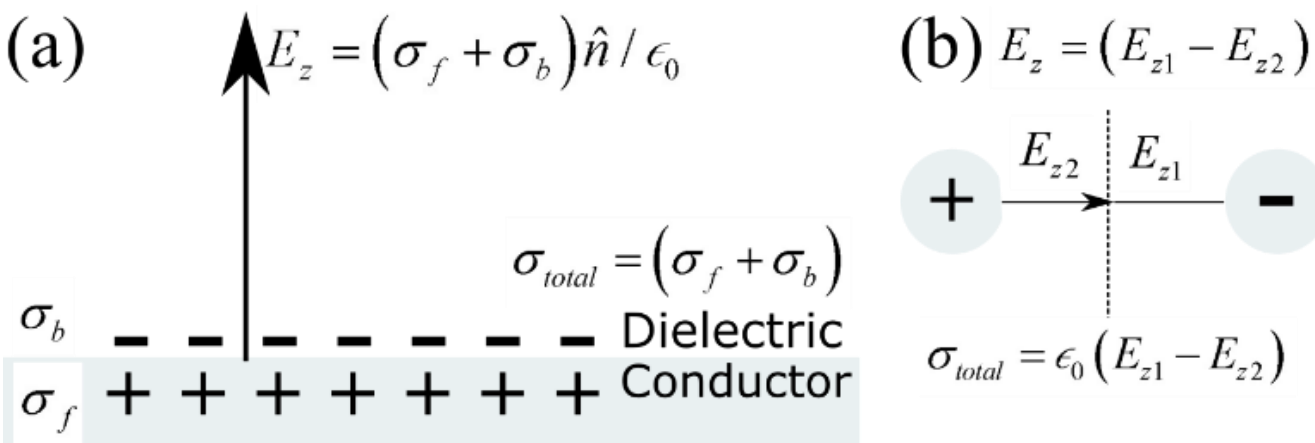

Figure A 1: (a) Relationship between the normal to the surface component of the electric field, $E_z$, and surface charges, where $\sigma_f$ is the free surface charge density inside the metal and $\sigma_b$ is the bound surface charge density in the dielectric. (b) Electric field line associated to a pair of opposite charges, crosses the interface normally, hence $Ez = E_{z1} - E_{z2}$.

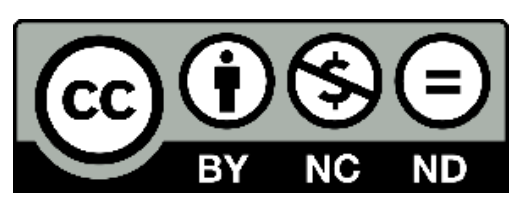